\documentclass[english,11pt]{article}

\usepackage{amssymb}
\usepackage{graphicx}
\usepackage{amsfonts}
\usepackage{bbm}
\usepackage[T1]{fontenc}
\usepackage[latin1]{inputenc}
\usepackage{babel}
\usepackage{color}
\usepackage{epsfig}

\begin{document}

\title{The Dark Universe Riddle}%

\author{ A. J. S. Capistrano\footnote{capistranoaj@unb.br}\\
Universidade de Bras\'{\i}lia, Instituto de F\'{\i}sica,
Bras\'{\i}lia, DF.70919-970\\
P.I. Odon\footnote{podon@maimonides.com}\\
Maimonides School, Boston MA, USA}

\maketitle

\begin{abstract}
In this work we review some of the theoretical efforts and
experimental evidences related to Dark matter and Dark energy
problems in the universe. These dilemmas show us how incomplete our
knowledge of gravity is, and how our concepts about the universe
must at least be revised. Mainly, on the Wilkinson Microwave
Anisotropy Probe (WMAP) fifth year, the data indicates that more
than 90$\%$ of the total energy density of the universe is dark.
Here we discuss the impact of these phenomena imprint on
gravitational and quantum field theory's standard history. Moreover,
we point out some recent and upcoming projects on Cosmology intended
to shed light on these problems.

\textbf{keywords}: \emph{Dark matter, Dark energy, Cosmological
Constant, alternative theories}.
\end{abstract}

\tableofcontents

\section{Introduction}
Our understanding of the universe has dramatically changed in the
past decades. Cosmology has actually become an experimental based
discipline with remarkable development and high level of confidence
when regarding measurements. Ever since the interest on cosmology
has grown, the scientific community began to realize that there is a
lot more to the universe than it meets the eye. Indeed, a
sought-after consistent physics theory must be a successful
combination between theoretical predictions and phenomenology, and
efforts from several research groups have been made in this
direction. Moreover, in recent decades, there were an amazing
development of physical theories, some of them beyond the standard
frameworks of \emph{Quantum field theory}(QFT) and \emph{General
relativity} (GR); due to the lack of a proper solution by the
mainstream theories, new theories have contributed to the most
intriguing and exciting scenarios physics has ever seen. For
instance, a vast new multidimensional world dominated by
superstrings and/or branes has been developed, and more recently,
Brane-worlds, became the ultimate quest to describe nature. This review aims at some capital problems in modern Astrophysics and
Cosmology concerning the dark matter and dark energy problems, and drawing some attention to the hierarchy problem between gravitation and gauge interactions. However, this note does not aim to be a complete account of
these problems and it is far from being due to its large influence and complexity, and we restrict ourselves to the discussion of the most important points in respect to some basics theoretical and phenomenological issues.

The paper is organized as follows: the second section briefly discusses the unification
of the fundamental interactions and the understanding of gravitation
within theories of spin-2. The underlying point is that the ``dark
problems'', as we call them, require an understanding on how gravity
interacts with strong, weak and electromagnetic fields in nature.
Moreover, the subsequent sections refer to the discussion of dark
problems {\it per se}. Therefore, the third section discusses
\emph{Dark matter} from its original motivation in the early 1930's,
with studies on galaxies and clusters of galaxies movements, all the
way to the importance of recent collected data and its effect on
unification theory candidates.

In addition, in the fourth section, we discuss the \emph{Dark
energy} problem \cite{perlriess,perlriess2} consisting in an
unclustered component of negative pressure related to the
accelerated expansion of the universe. In particular, this problem
includes in itself another problem discussed in this article: the
\emph{Cosmological Constant}. At first neglected by Einstein himself
as his Greatest blunder, the cosmological Constant has been regarded as an odd solution for
the dark energy problem. And in the last section, we make brief
comments on some recent and upcoming projects, whose main purpose is
to detect effects related to dark matter and dark energy in the
universe. The final comments are in the conclusion section. 

Hereafter, for the sake of notation, we use capital Latin indices
that run from 1 to 5, Greek indices are counted from 1 to 4 and
small case Latin indices run from 1 to 3 and the index 4 refers
specifically to the time coordinate. The colon and semicolon refer
to ordinary derivative and covariant derivative, respectively.

\section{Remarks on the quest for unification of the fundamental
interactions}\label{2}
\begin{flushright}
\begin{quote}
   \footnotesize{ \emph{Although Einstein's theory of the gravitational field is the
most widely accepted theory of gravitation, it is rather
disconcerting to note that Einstein's theory appears to be
strikingly different from the present theories of the
electromagnetic field and the meson fields}(...)} S.N.
Gupta\cite{gupta}
\end{quote}
\end{flushright}

After the success of GR in 1916, bringing depth to Newton's
gravitational theory, theoretical efforts to merge gravitation and
electromagnetism into a unique scheme of unification dramatically
increased over the following decades, after all that was the only
gauge interaction conceived at that time. Since gravity does not
match the gauge interactions as posed by the hierarchy problem of
the fundamental interactions, that is, the quantitative difference
between electroweak and Planck scales $\left(m_{pl}/m_{EW}\sim
10^{16}\right)$ based on the coupling constants measurements, we
find necessity in briefly revising some proposals in a manner to
realize how knowledge of gravity was developed and how strange its
behaviour can be mainly on the current dark matter and dark energy problems.


\subsection{Weyl and Kaluza-Klein theories}
To do so, we have to look back upon the first half of the XX
century. In the end of the 1910's, Weyl's theory \cite{weyl} was the
first attempt to unify gravity and electromagnetism. He altered
Riemann's geometry with a non-vanishing metric condition
\begin{equation}
g_{\mu\nu;\rho}=\;-2A_{\rho}g_{\mu\nu}\;\;,
\end{equation}
where $g_{\mu\nu}$ is the metric tensor and $A_{\rho}$ is the
4-vector electromagnetic potential. Hence, one can find the
connection
$$
\Gamma_{\nu\lambda}^{\mu} = \frac{1}{2}
g^{\mu\sigma}(g_{\lambda\sigma, \nu} + g_{\sigma\nu, \lambda} -
g_{\lambda\nu, \sigma}) + g_{\mu\sigma} (g_{\lambda\sigma} A_{\nu} +
g_{\sigma\nu} A_{\lambda} - g_{\lambda\nu} A_{\sigma})\;,$$ and
write the action
\begin{equation}\label{eq:weyl}
S=\frac{1}{2} \int d^4x \sqrt{-g} \left\{\frac{1}{2} F_{\mu\nu}
F^{\mu\nu} + (^\ast R)^2 \right\}\;,
\end{equation}
where $k_{\mu}$ is an arbitrary vector. The $F_{\mu\nu}$ tensor is
given by
\begin{equation}
F_{\mu\nu}= k_{\mu , \nu} - k_{\nu , \mu}\;,
\end{equation}
which obeys the contracted Bianchi identities and it can be related
to Maxwell tensor. In addition, the modified scalar curvature $^\ast
R$ is given by
\begin{equation}
^\ast R= R + 6k_{\; ;\mu}^{\;\mu}  - 6 k_{\mu}k^{\mu}   \;,
\end{equation}
where $R$ is the usual scalar curvature. Thus, Weyl built his theory
in a curved space-time geometry and due to GR,
Poincar$\acute{\mbox{e}}$'s symmetry from electromagnetism was
replaced by a diffeomorphism group of the space-time, which led to a
non-gauge invariance theory of coordinate transformations.
Essentially, it means that any gauge and any solution of the
equations are not valid to all observers, making it incompatible
with electromagnetism, which is based on firm experimental grounds.
For instance, in Weyl's theory, a bar moving under influence of the
electromagnetic field could have its length changed, which is
clearly incompatible with observations
\cite{schouten,O'Raifeartaigh}. In a geometrical sense, the parallel
transport in Weyl's geometry could modify both direction and the
length of a vector in this space.

Accordingly, still in 1919 (actually, the paper was published only
in 1921), motivated by the previous Weyl's work, Kaluza
\cite{kaluza} proposed a five-dimensional theory where
electromagnetism was compactified in an extra-dimension small circle
$S_1$ built at each point over the Minkowski space-time $M_4$. Actually, this was
not a new address, Nordstr$\ddot{\mbox{o}}$m
\cite{O'Raifeartaigh,nords} in 1914 had already made an attempt to
unify electromagnetism to gravity considering the ``cylinder''
condition which means that it can be easily expanded in Fourier
modes and proved by Einstein and Bergman proposal in which the metric of the \emph{bulk} $\mathcal{G}_{AB}$ was given by
$$\mathcal{G}_{AB}=\sum_{n}\mathcal{G}_{AB}^0(x) e^{\frac{in\pi\phi}{c}}\;,$$
where $\phi$ is the coordinate related to the extra dimension. Thus, for instance, let be a scalar field $\phi$ we can write the
periodic condition $\phi(x)=\phi(x + 2\pi R)$, where $R$ is the
radius of cylindrical extra-dimension. Hence, one can determinate
discrete $n$ values for the momenta as
\begin{equation}
  p=\frac{n}{R}\;,\;n=0,1,2...
\end{equation}
The set of states is currently called nowadays as the tower of
Kaluza-Klein modes. In addition, Kaluza used the topology $M_4\times
S_1$ of the 5-dimensional space-time with the \emph{ansatz}
\begin{equation}
\mathcal{G}_{AB}=\left(
                    \begin{array}{cc}
                      g_{\mu\nu} + A_{\mu}A_{\nu} & A_{\mu} \\
                      A_{\nu} & \phi \\
                    \end{array}
                  \right)\;\;,\end{equation}
where $\mathcal{G}_{AB}$ was the 5-dimensional Riemann metric,
 and $g_{\mu\nu}$ and $g_{ij}$ are components of spin-2 and
spin-1 particles respectively; both are described from the point of
view of a 4-D observer where the $g_{55}$ component can be regarded
as a non-massive scalar field $\phi$.

In 1926, Klein proposed that Kaluza's theory was only valid in a
quantum regime of order of  $10^{-33}$cm (the Planck length), which
corresponds to the Planck energy of $10^{19}$Gev in addition to the
condition of cylindricity besides stating that $g_{55}$ is a
constant with $\phi=1$. Despite the fact that the fifth dimension
provides an uneasy feeling, the Kaluza-Klein theory was very
suggesting because \emph{a priori} it unified gravity and
electromagnetism with the lagrangian
\begin{equation}
\mathcal{L}= R
\sqrt{\mbox{det}(g_{\mu\nu})}+\frac{1}{4}\;\mbox{tr}\;\left(F_{\mu\nu}F^{\mu\nu}\right)\;\;,
\end{equation}
where $F_{\mu\nu}$ is the Maxwell tensor. However, Klein's proposal
imposed serious constraints on the theory, which should predict a
massive and detectable particle according to the solution of
Klein-Gordon's equation, but it never happened \cite{duff}. Some
efforts for saving the theory were made in subsequent decades with
the Kaluza-Klein non-abelian approach.

In 1926-27, Fock \cite{fock} and London \cite{london} pointed out
that Weyl's theory could be relevant in the quantum context \emph{if
dissociated from gravity}. Thus, based on Weyl's theory, they
brought up the concept of a local \emph{gauge} transformation, e.g
$$\Psi'(x)= \exp{[i\xi(x)]}\; \Psi(x), \;\Psi'^{\ast}(x)= \exp{[-i\xi(x)]}\; \Psi^{\ast}(x)\;\;,$$
where $\Psi(x)$ and its conjugated $\Psi^{\ast}(x)$ are regarded as
complex functions of a wave field, and $\xi(x)$ is a phase
coordinate-dependent parameter. Even though the theory had seem to
be successfully retrieved, it came at odds with isospin
\cite{marateck} observations in the strong field scale. The isospin
presented a \emph{global} symmetry, called after $SU(2)$ symmetry
group, in contrast with the local gauge in Weyl's theory. Hence, the
internal gauge transformations gave a final strike on Weyl's theory
as a theory of unification.

\subsection{Gupta's and ADM scheme}
Another interesting approach was made in 1954 by Gupta \cite{gupta},
whose original intention was to study spin-2 fields. He proposed a
theorem which establishes that the spin-2 fields in a Minkowski
space-time can be described by an Einstein-type system of field
equations. His motivation was the study of a linear massless spin-2
field in the Minkowski space-time, a theory first conceptualized by
Pauli and Fierz \cite{pauli}. Gupta's new theorem was very
attractive because it showed a remarkable resemblance with the
linear approximations of Einstein's equations for the gravitational
field. In this sense, he linearized Einstein's equations in
Minkowski flat space-time with an infinite number of terms in the
Lagrangian density. Thus, the linearized Einstein's equations could
be written as
\begin{eqnarray}
 &&\epsilon^{\alpha\beta}\frac{\partial^2g^{\mu\nu}}{\partial x^{\alpha}\partial x^{\beta}}=\tau_0\Theta^{\mu\nu}\;,\\
&&\Theta_{\mu\nu;}^{\;\;\;\;\;\;\nu}=0\;,
\end{eqnarray}
where  $\tau_0$ is a constant and $\epsilon^{\alpha\beta}$ is a set
of quantities given by
\begin{equation}
\epsilon^{\alpha\beta}= \left(
                  \begin{array}{cccc}
                  -1 & 0 & 0 & 0 \\
                  0 & -1& 0 & 0 \\
                  0 & 0 & -1&  0\\
                  0 & 0 & 0 & +1 \\
                  \end{array}
                  \right)
\;\;.
\end{equation}
It is important to note that $\Theta_{\mu\nu}$ is the symmetrical
energy-momentum pseudotensor, where the supplementary condition
$g_{\mu\nu;}^{\;\;\;\;\;\;\nu}=0$ applies. A similar situation
occurs on Maxwell's theory
\begin{equation}
\Box^2A_{\mu}=-(1/c)j_{\mu}\;,
\end{equation}
 with the Lorentz gauge
\begin{equation}
A_{\mu\;, \mu}=0\;,
\end{equation}
where $A_{\mu}$ is the electromagnetic potential and $j_{\mu}$ is
the current four-vector. Thus, according to Gupta, the same
rationalization can be applied to a spin-2 field such that
\begin{eqnarray}
 &&\Box^2 U_{\mu\nu}=\tau_0 \Omega_{\mu\nu}\;,\\
&&U_{\mu\nu;}^{\;\;\;\;\;\;\nu}=0\;,\nonumber\\
&&\Omega_{\mu\nu;}^{\;\;\;\;\;\;\nu}=0\;,\nonumber
\end{eqnarray}
where  $U_{\mu\nu}$ is a real symmetrical tensor, $\tau_0$ is a
coupling constant and $\Omega_{\mu\nu}$ is a conserved symmetrical
tensor, which can also be written as $t_{\mu\nu}$. The $t_{\mu\nu}$
tensor represents the tensor for the gravitational field
energy-momentum plus $T_{\mu\nu}$ tensor, which represents the
tensor for the energy-momentum of the gauge interactions. The same
equation can also be derived from a variational principle, leading
to the Lagrangian density
\begin{equation}
\mathcal{L}=-\frac{1}{2}\frac{\partial U_{\mu\nu}}{\partial
x_{\lambda}} \frac{\partial U_{\mu\nu}}{\partial x_{\lambda}}+ f_1+
f_2+...\;\;\;,
\end{equation}
where the infinite terms $(f_1, f_2, ...)$ compose the
energy-momentum $t_{\mu\nu}$. Even further, if we consider only
gravitation, one can write a set of Einstein-type equations
\begin{equation}
U_{\mu\nu}-\frac{1}{2}Uu_{\mu\nu}=\alpha \tau_{\mu\nu}\;,
\end{equation}
where the symbols $u_{\mu\nu},U_{\mu\nu}$, and $U$ can be regarded
as a metric-type tensor, a Ricci-type tensor and a scalar-type
tensor respectively. Clearly, this new geometry was a copy of
Riemann's geometry with a metric and a curvature associated to it.

The shortcoming of this scheme is that Gupta assumed that the
physical quadridimensional spacetime was flat; that induced a
geometrical inconsistency by the metric tensors $g_{\mu\nu}$ and
$u_{\mu\nu}$. Moreover, in 1970 Deser \cite{deser} showed a
generalization of Gupta's theorem suggesting that it could be
possible to apply such mechanism to Yang-Mills' theory in a manner
to be derived from a similar argumentation. In 1978, Fronsdal
\cite{fronsdal} proposed a generalization of the theorem for
arbitrary \emph{spins} of massless fields. As far as we know, at
least over the last decade, there is no trace in literature of a
work based on such theorem, except in \cite{Francisco} which
made an approach to deal with strong gravity and spin-2 fields in
order to associate extrinsic curvature, which in differential
geometry is responsible for measure the divergence or convergence of
the normal vector with respect to the surface, to a fundamental
spin-2 field in nature.

Another interesting attempt was made in 1962 by Arnowitt, Deser and
Misner (ADM) \cite{adm}. The ADM theory was based on the attempt of
making a canonical quantization of gravitation in a manner to deal
with quantum fluctuations of 3-dimensional hypersurfaces. The
three-plus-one dimensional decomposition of the Einstein field leads
to the line element decomposition
\begin{equation}
    ds^2= -N^2dt^2 + \left( N^{i} dt + dx^{i}\right)\left( N^{j} dt +
    dx^{j}\right) g_{ij}\;,
\end{equation}
where the time component is given by
\begin{equation}
\bar{g}_{44}= N^{i}N^{j}g_{ij}- N^2\;,
\end{equation}
and
 \begin{equation}
\bar{g}_{4j}= N^{i}g_{ij}\;,\; \bar{g}^{\;44}= -(N)^{-2}\;,
 \end{equation}
 \begin{equation}
\bar{g}^{ij}= g_{ij} -
\frac{N^i}{N^2}N^{j}\;,\;\sqrt{\mbox{det}(\bar{g}_{\mu\nu})}=N\sqrt{\mbox{det}(g_{ij})}\;.
 \end{equation}
where the overbar indicates a four-dimensional quantity. The $N$ is
the lapse function and $N^{i}$ are the components of shift vector
field. They are Lagrange multipliers and determine, for instance,
the deformation of a three-dimensional space-type hypersurface
$\sigma$ at time $t$ to another hypersurface $\sigma'$ at time
$t+dt$ in a space time $M_{4}\times \sigma$. Moreover, from the
$(3+1)$-decomposition of the Einstein-Hilbert action
\begin{equation}
S= \int dt \int_{\sigma} dx^3 (\pi^{ij} \dot{g}_{ij}- N^{i}
\mathcal{H}_{i}- N \mathcal{H})\;,
\end{equation}
where the dot means time-derivative,
$\pi^{ij}=\sqrt{\mbox{det}(g_{ij})} (k^{ij} - g^{ij}k)$ and $k^{ij}$
is the extrinsic curvature projected on the $\sigma$ surface. Thus,
one can obtain the super-momentum and super-Hamiltonian constraints
on $g_{ij}$ and $\pi^{ij}$
\begin{equation}
    \mathcal{H}_{i}= -2 \pi^{j}_{\;\;i;j}=0 \;,
\end{equation}
and also
\begin{equation}
\mathcal{H}=
    \left(\mbox{det}(g_{ij})\right)^{-1/2}g_{ij}g_{kl}(\pi^{ik}\pi^{jl}-\frac{1}{2}\pi^{ij}\pi^{kl})-
\sqrt{\mbox{det}(g_{ij})}R= 0\;,
\end{equation}
where $R$ is the three-dimensional scalar curvature.

The formulation came to fail due to the arbitrary diffeomorphism
transformations, which imposed a constraint on the Poisson brackets
structure. Basically, the Poisson brackets do not propagate
covariantly which still remains as a keen obstacle to quantization
of gravity even in other theories. Otherwise, Dirac took into
account of this problem stating that the general covariance was the
main shortcoming for this method of trying quantization. According
to him, it was only possible when one chose normal direction of
propagation \cite{Dirac}. A good review of Dirac's method can be
found in ref.\cite{Matschull}. Recently, adapting to the brane-world
models, some authors \cite{Maia} demonstrated that quantization of
gravity is possible based on ADM formulation due to the confinement
of the diffeomorphism transformations, which do not ``leak'' into
the bulk, where the brane is embedded, but a lot of work is yet to
be developed in order to close this argument.

\subsection{Non-abelian Kaluza-Klein, two-tensor metric theories and other comments}
In addition, in 1965 \cite{O'Raifeartaigh} the non-abelian approach
to Kaluza-Klein's theory was developed. In this new theory, the
space-time was defined by the topology product $M_4\times B_N$,
where $B_N$ is a compact inner space. This geometry was the solution
to Einstein's equations on $(4+N)$-dimensions. The metric \emph{ansatz} was given
by
\begin{equation}
\mathcal{G}_{AB}=\left(
                    \begin{array}{cc}
                      g_{\mu\nu} + g_{ab}A^a_{\mu}A^b_{\nu} & A_{a\mu} \\
                      A_{\mu a} & g_{ab} \\
                    \end{array}
                  \right)\;\;,
\end{equation}
where $\mathcal{G}_{AB}$ is the Riemannian metric in N-dimensions
and $A^a_\mu$ is the Yang-Mills' potential, i.e, the connection
associated to a $SU(N)$ symmetry. The $g_{ab}$ component can be
regarded as a Killing's form of Lie's algebra for the gauge groups,
and $(a,b)$ indices are Lie indices of $SU(N)$ group. Hence, the
outcome lagrangian was
\begin{equation}
\mathcal{L}=R\sqrt{\mbox{det}(g_{\mu\nu})}+\frac{1}{4}\;\mbox{Tr}\;\left(F_{\mu\nu}F^{\mu\nu}\right)\;\;,
\end{equation}
where $F_{\mu\nu}=[[D_\mu, D_\nu], D_\mu]$,
$D_\mu=\mathbf{1}\partial_\mu+ g A_\mu$ and $g$ is a coupling
constant.

Although having a interesting structure, the non-abelian
Kaluza-Klein theory needed to be submitted to experiments. When
applied to the experimental observations, the theory fell through on
the fermionic chirality due to the prediction of a huge fermionic
mass which was at odds with the observed helicity at electroweak
scale. Besides, there were also inner theoretical inconsistencies
such as the size of the $B_N$ space and the definition of the
``ground state'' (flat, deSitter or Anti-deSitter) of the theory, as
pointed out by Abbot and Deser in 1982 \cite{abbot}. In a manner to
give a proper solution to the fundamental problem, in 1983 Rubakov
and Shaposhnikov \cite{ruba} proposed a bidimensional model where
the gauge interactions would be confined under a potential well.
They did not succeed. The $B_N$ space was not observed because it
had a different scale of order of Planck scale which created another
observational limitation. Nevertheless, this original work can be
regarded as the former inspiration for the current brane-world
models. The theory endured untill 1984, when observational
inconsistences were discovered, even so, several studies based on
Kaluza-Klein theory and its reinterpretation have been made in
several works \cite{elgaroy,elgaroy2,elgaroy3,elgaroy4}.

In 1970, based on Gupta's theorem and the photon-meson-$\rho$ model,
Isham, Salam and Strathdee \cite{Salam} idealized a theory in which
the spin-2 field was an effective field for short distances. It was
assumed that the existence of a new tensor field $f_{\mu\nu}$ or a
$f$-field that should describe a spin-2 massive particle called
\emph{f-meson} would couple directly to the hadronic matter. The
hadrons and leptons were regarded as high and low energy interacting
particles, respectively. Moreover, Einstein's graviton $g$-field
would describe leptons that couple to hadronic matter only through a
\emph{f-g} mixing term. As an application of Gupta's theorem, the
\emph{f-g} theory was built on the 4-dimensional Minkowski
space-time.

Criticizing this scheme, in 1973 Aichelburg \cite{aichelburg} stated
that it was impossible to build a theory with two metric tensors in
the same space-time without losing causality. On their defense, some
authors \cite{dirac,rosen} assured that to make an unified theory,
the adoption of two metric tensors is required to deal with atomic
and gravitational phenomena. However, it was proven that the
\emph{f}-meson was a resonance of a quark bound state with short
lifetime deprived from a fundamental meaning. Nonetheless, in the
same year, Dirac \cite{dirac} proposed another theory using two
metrics. He retrieved Weyl's theory by adapting it to his hypothesis
of Large numbers, originally proposed in 1938 \cite{dirac2}.

Dirac's large numbers proposal attributed a time-dependence to the
gravitational constant, such that $G\approx \frac{1}{T}$ and $T$ is
the age of the universe differently from GR, where, hereafter, the
gravitational constant $G$ has a fixed \emph{three-dimensional}
value in accordance with the Newtonian theory. Moreover, with the
same definitions of eq.\ref{eq:weyl}, Dirac added a scalar field to
Wely's action and found the following functional
\begin{equation}
S[\phi, k_{\mu}]=\frac{1}{2} \int d^4x \sqrt{-g} \left\{\frac{1}{2}
F_{\mu\nu} F^{\mu\nu} + ^\ast R \phi^2 + \alpha
\phi_{\ast\mu}\phi^{\ast\mu}\right\}\;,
\end{equation}
where $\alpha$ is a dimensionless constant. Thus, the original
Weyl's theory was modified by replacing the term $\left(^\ast
R\right)^2$ by $^\ast R\phi^2$ \cite{dirac,mirabotalebi}. The
$\phi_{\ast\mu}$ term is the co-covariant derivative of $\phi$ and
is defined as
\begin{equation}
\phi_{\ast\mu}=  \phi_{\mu} + \phi k_{\mu}\;.
\end{equation}
For many authors, the large numbers are only the result of
numerological coincidence, until now without any experimental
indications. Even so, such hypothesis is still considered by some
authors \cite{sidhart,sidhart2,sidhart3} mainly of the strong
influence of the cosmological constant problem and hierarchy problem
of the fundamental interactions have imprinted in the recent years.

In addition, still in 1973, besides of Dirac-Weyl's and \emph{f-g}
theories, Rosen \cite{rosen} proposed also the \emph{bimetric}
theory. This theory was built on Minkowski flat space-time with two
symmetric tensors. The so-called $\Gamma_{\mu\nu}$ tensor would
describe properties of space-time and was interpreted as a second
rank tensor of spin-2. The $g_{\mu\nu}$ tensor was interpreted as a
gravitational potential tensor and was responsible for making the
interaction between gravitation, matter and other fields. As a
criticism to GR, Rosen's bimetric theory does not provide such
singularities as \emph{black-holes} and, moreover, it can provide a
gravitational energy momentum-tensor. Hence, when applied to
cosmology, the universe predicted does not have an initial
singularity as the \emph{big-bang} being closed in space (closed
curvature; see subsection \ref{ecosmo}) and eternal in time
\cite{rosen2, Francisco}. The main shortcoming of the theory is that
provides a dipole gravitational waves, instead of a quadrupole modes
like GR, which is at odds with the binary pulsar PSR1913+16
measurements. As well known, the binary pulsar PSR1913+16 has proven
to be a valuable tool to test alternative gravitational theories
\cite{will}. This fact made the theory lose any theoretical interest
until now.

On the other hand, it is instructive to point out that between 1930
and 1970, there was a great development on the unification process
dissociated from gravitation. The main contestant was the Yang-Mills
scheme. In spite of some difficulties, as how to allocate all the
particle groups in families, which would require an adoption of
other sort of symmetries in a Grand Unified Theory (GUT) proposal,
the standard model of gauge interactions unified electromagnetism,
weak and strong interactions in the group $U(1)\times SU(2)\times
SU(3)$.

In summary, the coupling of gravity to other fields constitutes a
hard task due to the lack of understanding of what gravity really
is. The problem persists because we still do not have a definite
quantum gravity theory, despite of the advent of M-theory and
brane-worlds. As stated by Misner \cite{misner}, gravity does not
behave as a gauge theory, that is, there is a qualitative issue that
makes gravity different from other gauge interactions. As we are
going to show throughout the next sections, dark matter and dark
energy problems, essentially, as far as we conceived these days,
both being effects of gravitation, aggravate this difference and require a deeper insight on the
meaning of gravity and how it interacts with others fields constituting an additional barrier to an effective unified field theory.

\section{Dark Matter phenomenology}\label{3}
\subsection{When the universe became ``dark''}
Our standard knowledge of gravitation, formation and evolution of
the universe is based on \emph{General relativity}. It has a
fundamental assumption that describes the universe in a reasonable
manner: the so-called \emph{Cosmological principle}. In short, the
cosmological principle states that in large scales, based on
redshift surveys up to 100Mpc (1Mpc$\sim 3,08\times10^{24}cm$) and the measurement of the galaxy 2-point correlation function \cite{totsuji}, the
observable universe (approximately 3000Mpc) can be regarded as
homogeneous and isotropic. However, considering cosmological
distances lesser than 100Mpc, inhomogeneity takes place. On this
scale, galaxies, cluster and superclusters of galaxies have greater importance,
which the Newtonian theory supposedly could be applied with a
reasonable level of confidence, completing studies of kinematics and
dynamics of these objects.

On the other hand, since the beginning of the 20th century, there
has been observed some astrophysical problems regarding rotation of
galaxies and clusters of galaxies. For instance, in the end of the
1920's by Oort \cite{bertolami} who pointed out the differential
rotation of the Milk-way, that is, the velocity in the core of
galaxy was bigger than the velocity in its outskirts. In subsequent
studies, Oort \cite{oort} noted that in the outer parts of galaxies,
stars were moving faster than predicted by Newtonian gravity,
suggesting that an additional force should exist to maintain stars
orbiting one galaxy. Actually, what is really observed is the velocity of regions of hydrogen clouds which varies on the distance in respect to the core of the galaxy.

A similar situation appeared to occur at cluster scales. In 1933, F.
Zwicky \cite{Zwicky} noted that the galaxy velocities in the COMA
cluster of galaxies were also moving faster than the velocities
predicted by Newton's gravitational theory. The velocity of the
cluster, as well as its stability could not be justified by only
taking into account its visible mass. He found a total mass
approximately 400 times bigger than the expected, considering the
number of galaxies and clusters' luminosity. Zwicky named it the
\emph{missing mass problem}.

Surprisingly, the problem pointed out by Zwicky was almost forgotten
by the scientific community until the 1970's, when it was retrieved
by Rubin, Ford and collaborators
\cite{Rubin,Rubin1,Rubin2,Rubin3,Bosma}, who obtained experimental
evidences to support Zwicky's observations. They were studying the
path described by stars in galaxies where the function between the
velocity of a star and its distance from the center of the galaxy is
usually called the \emph{rotation curve}. The studies of rotation
curves play an important contribution on our understanding of the
formation of the galaxies, particularly in spiral galaxies where the
galactic disks are observed \cite{Rubin3,Bosma,mo}. For short, they
discovered an unusual high speed of stars in the edge of spiral
galaxies, completely contradicting Kepler's theory, where a slow
down scenario is expected. Thus, in the rotation curves problem,
\emph{we should also add mass to maintain the galaxy's stability},
in the same way that it should be done to galaxies' clusters. When
we study such motions, away from the core of galaxy, it is verified
a discrepancy between the observed velocity and the theoretical
prediction. Moreover, the same anomaly also appears in the
elliptical galaxies case \cite{Fabricant,Fabian,Quinn,Hernquist}.
It is important to note that Zwicky's proposal runs in the Newtonian
context. In other words, Newton's theory should hold true also in
galactic scales. Hence, the solution for the missing mass problem
came with the idea of adding ordinary baryonic mass to the systems
studied as a manner to preserve Newton's theory. We can understand
baryonic matter as the common matter composed of elementary
particles(quarks and leptons) from the standard model of particle physics.

Another intriguing effect is the Pioneer anomaly \cite{nobb,bertol}.
The two space probes, Pioneer 10, launched on March 2nd 1972 to
visit Jupiter, and Pioneer 11, launched on April 5th 1973 to Saturn,
which they are under influence of an \emph{unexplained} constant
acceleration directed towards the sun with constant approximate
value of $a= (8\pm 1.3)\times 10^{-10}m.s^{-2}$. Surprisingly
enough, this phenomenon is compatible with the numerical value of
acceleration constant of the MOdified Newtonian Dynamics (MOND)
\cite{Milgrom1,Milgrom2}, proposed by Milgrom in the beginning of
the 1980's. In principle, this deviation should had been explained
by a gravitational field produced by GR, but it was not. Among other
explanations for the Pioneer phenomenon, \emph{a local effect, at
solar system level} of dark matter also has been proposed to explain
such anomalous behaviour of the space probes \cite{nieto} at odds
with other works which state that the dark matter solar system is
not related to Pioneer anomaly \cite{siegel} at all. However, with
the improvement of observational devices, dark matter reveals a much
more complex problem, decisively affecting the way we perceive the
universe.

\subsection{Can we understand Dark Matter?}
Our sense about dark matter changes as experiments become more and
more effective. As discussed before, according to Zwicky's
observations, dark matter consists of a sort of matter which neither
absorb, nor emit light, or any electromagnetic radiation in any
frequency bandwidth whatsoever. Its presence is revealed only
through gravitational field. The main mechanism to try to identify
dark matter concerns on gravitational micro-lensing effects. These
effects play an important role on mapping and measuring subtle
luminosity distortions of objects on the space-time background. This
can only be achieved through usage of advanced spectrographical and
optical telescopes, such as NASA's Chandra X-ray Observatory
\cite{chandra}, and Canadian-France-Hawaii Telescope (CFHT)
\cite{cfht} in Hawaii.

Recently, two merging phenomena in giant clusters gained substantial
attention for providing the first sought-after apparently
\emph{direct} evidence of dark matter based on weak lensing, x-ray
and visible optics astronomy. These measurements were collected by
the Chandra X-ray Observatory on the so-called \emph{Bullet cluster}
\cite{clowe} 1E0657-558, which essentially consists of two clusters
forming a bullet-like structure tied to galaxies but moving through
the intercluster plasma. Specifically, the center of mass of two
spherically symmetric dark matter halo does not match the center of
mass obtained by alterations of the gravitational force law with
respect to the Newtonian theory.

On the other hand, the Abell 520 cluster (MS0451+02) observed a dark
core \cite{mahdavi}, according to observations of CFHT, where dark
matter does not appear to be anchored to any other galaxy, but now
to the intercluster plasma. Note that we presented two situations
and two different outcomes. The silver line is pinpointing under
what conditions and constraints dark matter justify such anomalous
behaviours as those mentioned above. Nonetheless, due to the lack of
a proper explanation, one can appeal that these evidences indicate
an odd existence of dark matter, whatever that is \cite{govert}.

\subsubsection{Dark matter candidates}
In a manner to try to understand the nature of dark matter, two main
approaches have been considered: first, the \emph{Hot Dark Matter}
(HDM) model, which states that when the galaxies were first formed,
dark matter was composed of relativistic particles ($kT>>mc^2$).
This odd reference to ``temperature'' refers to the energy levels of
these particles, and to how fast they travel. The main candidates
are neutrinos and the hypothetic Strongly Interactive Massive
Particle (SiMPs) \cite{wandelt,teplitz}. But the quantity of
neutrinos \cite{Lesgo} required is larger than the observed, so they
can not lead alone to a large scale structure formation. Moreover,
relativistic particles do not clump, due to their high speeds, which
is at odds with observations about the universe structure. In addition, the
\emph{Weakly Interacting Massive Particles} (WIMPs) \cite{bednyakov} are another candidate for hot dark matter
which have as main candidates the supersymmetric particles called
\emph{neutralinos}. Until now, these constraints on HDM kept these hypotheses away as an odd
solution for the dark matter problem.

The second approach is the \emph{Cold Dark Matter} (CDM) hypothesis,
which consists of non-relativistic particles, such as the \emph{Massive Compact Halo Objects} (MACHOs).
MACHOs are just generic denominations to massive cosmic bodies,
such as planets as massive as Jupiter, or/and distributed in a
spherical halo, orbiting the galaxy itself, far away from the stars.
It is important to point out that MACHOs are composed of baryonic
matter. Moreover, the quantities required to solve the rotational
curve problem, or the large structure formation, are much bigger
than the observations of MACHOS in the universe. These components
have been observed with help of gravitational micro-lensing effects,
but only in very small amounts, far beyond the required quantity.

Another candidate to CDM is the Axion \cite{khalil}, which is a
hypothetic spin-0 particle, originally postulated by Peccei-Quim's
theory to solve CP problems in quantum chromodynamics. It was
regarded as a candidate because it has a very small mass $\left(10^6
- 10^2\;eV.c^{-2}\right)$ and zero electric charge, which is in cue
with why it can only be detected through gravitational effects.
However, current debates on the influences of dark matter on large
structure formation suggest that cold dark matter is related to a
gas of generical Wimps, due to some of dark matter's properties as
long-lived and stable particles, and regarded as the main character
to large scale formation issues.

\subsubsection{Dark matter on cosmological scales}
Today's sophisticated astrophysical experiments tell us that dark
matter cannot be understood only taking into account the local
rotational curves problems or slow moving objects in clusters and
colliding clusters. If one asks for the origins of dark matter, we
end up in the early universe, when dark matter was supposed to break
the cosmological homogeneity of baryons, thus creating large
structures, as observed today. In this context, the natural starting
point to study dark matter is its gravitational field.

In the cosmological scale, dark matter seems to be consistent with
the standard FLRW model, but only recently the cosmic microwave
radiation data collected by WMAP indicated that most of dark matter
content must be cold, and must be of \emph{non-baryonic nature},
 i.e, out of the standard model of particle physics. As suggested in some
works \cite{khalil,ostriker,mira}, cold dark matter plays a serious
role in large scale structures formation and could induce to a small
overdensity in the primordial universe after the initial inflation,
which could give birth to an extra gravitational attractive force,
leading to a large scale structure formation by growth of
perturbations and gravitational instability. According to WMAP data,
dark matter is a dominant component in the early universe
representing more than 60$\%$ of the total energy density.

As it happens, dark matter seems to be the ``spark'' that unrolled
the formation of large structures. Otherwise, if we consider the HDM
hypothesis, it produces a ``top-down'' scenario, where large
structure objects (clusters and superclusters)
 are formed before small structures, due to the
fact that relativistic particles do not clump. Nevertheless, there
are strong evidences of the existence of young galaxies
\cite{planck} of order of 500 or more celestial objects, which were
formed at least 13 thousand million years after the big bang
\cite{gamov}. This is a stunning fact, because according to the
big-bang theory, they would not have enough time to be formed. This
notion is at odds with galaxies evolutionary timescales. Other works
state that a Giant Boson star \cite{jae} could be a more reliable
model that provide a solution to, for instance, the flatness of the
rotation curves and halo formation to the detriment of the CDM
model. Therefore, either some exotic particles must be considered,
or else an adequate gravitational theory should be devised.

On the fifth year of WMAP observations revealed a new distribution
of the composition of the universe \cite{spergel,map}, which shows
that the thermal radiation is about $2.73 K$; also no substantial
amount of anti-matter was found. It also revealed that 23$\%$ of the
universe is composed of Dark Matter, 72$\%$ of Dark Energy and only
4.6$\%$ of visible or baryonic matter (H$\sim$ 75$\%$, He$\sim$
25$\%$ and trace amounts of heavy elements). Moreover, the 23$\%$
total energy density of the universe related to Dark matter is
mostly of cold (non-relativistic) and non-baryonic nature.
More specifically, the analysis of the power spectrum indicates that
a theory of gravity based essentially on the properties of baryonic
matter would produce a lower third peak \cite{spergel,nolta} and it
would be incompatible with the universe formation.
These evidences contributed to the acceptance of the CDM models to
the structure formation. Today, the most accepted phenomenological
models are also based on CDM proposal, for instance, the $\Lambda$
and X-CDM models, which will be discussed in section \ref{4}. For
further information see \cite{ostriker,gawiser,tsagas,pololo}.

\subsubsection{Alternatives to Dark matter}
The most simple model of dark matter consists on the so-called
\emph{Dark Matter Halo hypothesis}, in which a galaxy would be
embedded in a dark matter bulk which is extensively used in
simulations of dynamics of universe. Truth of the matter is, we are
not able to determine the extension of the dark matter bulk because
of technical detection difficulties, but we cannot quite turn our
faces from it because it seems to have some physical reality to it
\cite{salucci}. This concept of a dark matter bulk has its origins
in the end of 20's and mid 30's due to Oort and Zwicky's
observations. Hence, it was entirely based on Newton's theory
satisfying a particular symmetry and boundary conditions. The term
bulk here is used only to try to explain where the missing mass
would be allocated or how further it is extended, being a proposal
of how to recover a gravitational pull that should exist.
Since dark matter interacts with ordinary matter essentially by
gravity, and that is assumed to be of Newtonian nature, most present
dark matter models depend upon Newton's theory and, therefore, are
gauged by Newtonian gravity paradigms. Thus, a previously defined
gravitational theory must be postulated before the analysis of CMBR
power spectrum experimental results can be more conclusive. Indeed,
all simulations and comparisons are estimated with respect to
Newtonian forces derived from the Newtonian gravitational potential.
Even so, the recent data suggests that dark matter, whatever it may
be, induces serious constraints to gravitational theories based only
on baryons or in other words the inertia concept like MOND.

MOND \cite{Milgrom1,Milgrom2} has received a substantial attention
in recent years and it has been backed by a theory in which
Poisson's equation for the Newtonian gravitational field is replaced
by an equation like
\begin{equation}
<\nabla,\mu\left(\frac{|\nabla\Phi|}{a_0}\right)>=4\pi G\rho\;\;,
\end{equation}
where $\mu\left(\frac{|\nabla\Phi|}{a_0}\right)$ is a function to be
adjusted to the specific type of galaxy, $a_0$ is an acceleration
constant with magnitude $a\sim 9\times 10^{-10}m.s^{-2}$, $\Phi$ is
the Newtonian gravitational field and $\rho$ is the energy density
of the baryonic matter source. Recently, a relativistic theory of
MOND called TensorVectorScalar or TeVeS \cite{bekenstein,dodelson}
has been developed. This relativistic model includes tensors,
vectors and scalar fields in a manner of providing an alternative
cosmology and in some sense generalizing the original MOND to
cosmological scales.

On the other hand, it seems obvious that General Relativity regarded
as the correction to Newton's theory would be a natural candidate to
deal with the curve rotation problem and the dark matter problem.
The standard argument against the effectiveness of GR is that in the
core of galaxy gravitation is much stronger than in its external
points and that is where GR would hold and should provide the
required correction. However, it agrees with it precisely where
Newton's theory holds. This is due to the huge concentration of mass
in the core of the galaxy and it produces a spherically
gravitational field but beyond that region, the gravitational field
becomes sufficiently weak to be taken over again by its Newtonian
limit. As well known, Schwarzschild's solution is an exact spherical
solution of Einstein's equations and with correct assumptions we can
derive Newton's gravitational potential. Therefore, the
gravitational field should be Newtonian everywhere else in the
galaxy.

If we want to look further, we can extend our analysis to \emph{the
Parametric Newtonian Approximation} of GR, and find a gravitational
potential which decays by the law $\left(1/r^{3}\right)$
\cite{silva}. Unfortunately this law is not consistent with
observations, leading to a rapid decay of the rotational curve away
from the core. This rationalization implies that the exact solution
to Einstein's equations can be disregarded under dark matter
context. This is due to that the lack a proper justification why
general relativity cannot be used in the dark matter problem. For
instance, we present some argumentation: first, Newton's theory does
not describe a strong gravitational field like those observed at the
galaxy cores \cite{richstone,Lauer}. Secondly, the weak
gravitational lensing used to detect the presence of dark matter in
clusters cannot be described with Newtonian gravity. These two
evidences suggest that if the predominant gravitational field in
galaxies and clusters is due to dark matter, then the dismissal of
general relativity in favor of Newtonian gravity is not completely
justified.

These attempts to explain dark matter have motivated the emergence
of many others gravitational theories, like, for instance: (1)
Adding a scalar field to Einstein's equation, in such a way that the
scalar-tensor theory corrects the Newtonian limit \cite{fay}; (2)
Modifying the concept of time in general relativity, so that the
Newtonian limit of the theory differs from the original Newton's
theory \cite{behar,harnett}; (3) Adding a cosmological constant with
appropriate sign \cite{white}; (4) Including higher order curvature
terms in the gravitational variational principle \cite{capo}; (5) quantum cosmology based on \emph{a priori} stochastic process considering the universe as a pre-geometric system \cite{cahill};(6)
Several brane-world models and variants have been considered in the
hope that more general brane-world equations of motion may provide
the correct velocity curves
\cite{Mak,Vollick,Okada,Nihei,ichiki,lidsay}. Nevertheless, the
problem with the constraints provided by observations are still too
subtle and difficult to deal with. From the theoretical point of
view, in a self-consistent manner capable of taking into account
both cosmological and local effects of dark matter, a sought-out
dark matter model is still an ideal.

\section{The Dark Energy problem}\label{4}
In order to obtain more insight on the dark energy problem, it is
instructive to attain some remarks on its history before we discuss
the problem itself. Nowadays, the current theoretical debate about
 dark energy problem is related, \emph{mainly}, to the
cosmological constant, as we are going to point out in subsection
\ref{decc}.

\subsection{Einstein's dilemma and the cosmological standard
model}\label{ecosmo} After proposing his equations in the end of
1915, Einstein concluded that they could provide a non-permanent or
static universe. But at that time the universe was bounded by
western philosophical thoughts \cite{weeler}, which stated the
eternity of the universe without beginning or end.

In accordance with this philosophy and the lack of any precise
experimental data about the conditions of the universe, Einstein was
compelled to modify his original theory by introducing a new term
$\Lambda > 0$ in order to obtain static solutions. In 1917, Einstein
proposed that the equations could be written as
\begin{equation}
R_{\mu\nu}-\frac{1}{2}Rg_{\mu\nu}-\Lambda g_{\mu\nu}=-8\pi G
T_{\mu\nu}\;,
\end{equation}
by using the spherical metric  $ds^2=dt^2-R^2(d\chi^2+\sin^2\chi
d\theta^2+ \sin^2\chi \sin^2\theta d\phi^2)$, where $R$ is the
constant radius of a 3-D sphere, $\chi$ runs from $0$ to $\pi$, and
$c=1$. Hence, one could generate a static dust-dominated universe,
visualized as a perfect fluid with constant density $\rho= \Lambda
(8\pi G)^{-1}$, radius $r=(8\pi G \rho)^{-1/2}$ and mass $M=2\pi^3
r^3 \rho= \frac{\pi}{4}\Lambda^{-1/2}G^{-1}$.

Starting from these seminal results, Einstein stated that the
inertia of a body could be ``induced by its mass but not determined
by it'' \cite{bassalo}, in accordance with Mach's principle. This
principle was originally proposed by Mach \cite{mach} in 1883, which
consisted in a relativity of concepts of inertia at odds with the
Newtonian concept of absolute space and time. According to Mach, the
inertia of a body was generated by the influence of the entire mass
of the universe on the body. In fact, despite the fact that GR does
not fulfill all the requirements of Mach's principle, due to the
equivalence principle of GR, even so Einstein believed that by
introducing a cosmological term he would be able to solve this
question \cite{peebles}.

Still in 1917, de Sitter presented a new result by adding the
cosmological constant to Einstein's equation in vacuum
($T_{\mu\nu}=0$) with the line element
\begin{equation}
 ds^2=\frac{1}{\cosh^2Hr}[dt^2-dr^2-H^{-2}\tanh^2Hr
 (d\theta^2+\sin^2\theta d\phi^2)]\;,
\end{equation}
where $H=\sqrt{\Lambda/3}$ in a spherical quasi-static universe with
radius $R=3 \Lambda^{-1}$. Weyl \cite{weyl2} and Eddington
\cite{eddington} checked independently that in the de Sitter's
universe, two arbitrary test particles could repel each other. This
fact was the first theoretical evidence of a possible expanding
universe \cite{bassalo}.

In 1922, Cartan \cite{cartan} demonstrated that the most general
expression of Einstein's tensor was guaranteed by adding a term
multiplied by the metric, in accordance with Bianchi's identities.
Hence, the existence of $\Lambda$ is a consequence of the
``imprecision'' of Riemann's geometry with respect to the shape of
the objects. Thus, the dismissal of $\Lambda$ is only justified by
one of the following arguments: symmetry, or a observational data
constraint. Nevertheless, the proposal of a static or quasi-static
universes started to fail in the subsequent periods.

The first strike on the cosmological term came along in the same
year by Friedmann. Friedmann \cite{friedmann} published a paper in
which he demonstrated a dynamical solution \emph{without} the
cosmological constant by assuming a homogeneous and isotropic
universe. The element line proposed was
\begin{equation}
ds^2=dt^2-a^2(t)\left[\frac{dr^2}{1-kr^2}+r^2(d\theta^2+\sin^2\theta
d\phi^2)\right]\;,
\end{equation}
where $a(t)$ is \emph{a priori} an unknown function of time and $k$
is a constant. When applied to cosmology, $a(t)$ is the scale factor
which can describe the distance between comoving observers as a
function of time, and $k=0,\pm 1$, which corresponds to the spatial
curvature of the universe . Hence, the model predicts three
possibilities for the geometry of the universe: $k=0, -1$ or $+1$,
which corresponds to a flat (asymptotically expansion), parabolic
(contracting universe) and hyperbolic (eternal expansion) universe,
respectively which depends on the total mass of the universe. The
same results were obtained independently by A. Walker \cite{walker}
and H. Robertson \cite{robertson}, and with contributions of G.
Lema$\hat{\mbox{i}}$tre \cite{lemaitre}. It turned out to be the
well-known Friedmann-Lema$\hat{\mbox{i}}$tre-Robertson-Walker (FLRW)
metric.

It is important to point out that the Lema$\hat{\mbox{i}}$tre model
of the universe consisted of an intermediate solution. It began in a
static Einstein universe and led to a vacuum solution of the
expanding deSitter universe. The universe was originated by what he
named the \emph{primeval atom}, launching a primitive idea of what
we now know as the \emph{big-bang} model.

Before we proceed further, it is instructive to attain some aspects
of the FLRW model. First, Friedmann equation can be reproduced by
taking the energy-momentum tensor of a perfect fluid in comoving
coordinates
\begin{equation}\label{eq:fluid}
T_{\mu\nu}=(p+\rho)U_{\mu}U_{\nu}+p\;g_{\mu\nu},\;\;\;U_{\mu}=\delta_{\mu}^{4}\;,
\end{equation}
where $U_{\mu}$ is the 4-velocity, $\rho$ is the total density of
all matter-energy contribution and $p$ is the pressure of the
perfect fluid. Thus, using the local conservation law $T_{\mu
4;\;\mu}=0$, we obtain
\begin{equation}\label{eq:conserv}
\dot{\rho}=-3\frac{\dot{a}}{a}(\rho+ p)\;,
\end{equation}
where the dot represents time derivative. Secondly, by solving
Einstein equations with the FLRW metric, we can find the spatial
components the Raychaudhury acceleration equation taking into
account the cosmological constant $\Lambda$
\begin{equation}\label{eq:ray}
\frac{\ddot{a}}{a}=-\frac{4}{3}\pi G(\rho+3p)+ \frac{\Lambda}{3}\;.
\end{equation}
Therefore, from eq.(\ref{eq:conserv}) and eq.(\ref{eq:ray}), we can
obtain the Friedmann equation
\begin{equation}\label{eq:friedman}
\left(\frac{\dot{a}}{a}\right)^2+\frac{k}{a}=-\frac{8\pi G}{3}\rho+
\frac{\Lambda}{3}\;,
\end{equation}
where $a(t)$ is the scale factor of the universe. The geometrical
meaning of $k$ as the spatial curvature is given by the time
component $(44)$ of the Einstein equations. The Friedmann equation
describes the dynamics of the universe and its validity at all
times.

Alternatively, according to the experimental observations at
present, one can write the equivalent form
\begin{equation}
H^2=\left(\frac{\dot{a}}{a}\right)^2=
H_0^2\;[\Omega_R\left(\frac{a_0}{a}\right)^4+
\Omega_{M}\left(\frac{a_0}{a}\right)^3
+\Omega_{\Lambda}+\Omega_k\left(\frac{a_0}{a}\right)^2]\;,
\end{equation}
or in terms of redshift $z$
\begin{equation}
H^2=H_0^2\left[\Omega_{R}(1+z)^4 + \Omega_{M}(1+z)^3 +
\Omega_{\Lambda}+ \Omega_{k}(1+z)^2 \right]\;,\;
\frac{a_0}{a}=1+z\;,
\end{equation}
where $H_0=\left(\frac{\dot{a}}{a}\right)_0= 75 \pm
10\;km.s^{-1}.Mpc^{-1}$ is the present value of Hubble parameter,
i.e, the rate of expansion of the universe at present. The
cosmological parameter $\Omega$ is defined as $\Omega=
\rho(\rho_{crit})^{-1}$, where $\rho$ is the energy density and
$\rho_{crit}= 3H^2 (8\pi G)^{-1} \sim 10^{-29} g.cm^{-3}$. One can
think of the critical density as the minimal scape velocity when we
calculate the rocket problem in mechanics. However, in cosmology,
the critical density is regarded as the minimal amount of energy
density to maintain a homogeneous and isotropic universe. Hence, if
$\rho> \rho_{crit}$, or $\Omega > 1$, then $k=+1$, which gives a
closed universe; on the other hand, $\rho<\rho_{crit}$, or $\Omega <
1$, implies $k=-1$, an open universe, and finally, when $\rho=
\rho_{crit}$, or $\Omega = 1$, we have a flat universe $k=0$.
According to CBMR observations, the \emph{total} cosmological
parameter $\Omega$, or simply, $\Omega_{\;tot}$, varies as $0.98
\leq \Omega_{\;tot} \leq 1.08$ which suggest that we live in an
approximately flat universe \cite{padman,mukhanov}.

Considering all contributions to the content of the universe, we can
normalize the equation to one and obtain
\begin{equation}
\Omega_{\;tot}=\Omega_R+\Omega_B+\Omega_{CDM}+\Omega_{\Lambda}= 1.
\end{equation}

As it happens, $\Omega_R= 8\pi G \rho_r(3H^2)^{-1}$ corresponds to
the radiation contribution of the universe of order of
$5\times10^{-5}$ with energy density $\rho_r \sim a^{-4}$; and
$\Omega_{M}= 8\pi G \rho_m (3H^2)^{-1}$, related to the rest mass
density $\rho_m\sim a^{-3}$, is the content of non-relativistic
matter which we have separated into two parts: baryonic matter
$\Omega_B\sim 0.046$, and the cold dark matter $\Omega_{CDM}\sim
0.23$. If we reconsider the cosmological constant and impose that it
plays the role of dark energy, we have its contribution as
$\Omega_{\Lambda}= 8\pi G \rho_{\Lambda} (3H^2)^{-1}\sim 0.72$ and
$\Omega_k= 8\pi G \rho_k (3H^2)\sim 0.0\pm 0.1$, which represents
the contribution of the geometry or curvature of the universe with
energy density $\rho_k\sim a^{-2}$.

In addition, we point out some experimental facts which contributed
decisively for the dismissal of a cosmological term at that time.
For instance, the measurements of redshift of galaxies by Slipher
\cite{eddington} in 1924, as well as Hubble's pioneer observations
of homogeneity and isotropy \cite{hubble1} in large scales and
redshift \cite{hubble2} in 1929. It all provided acceptance of the
FLRW model. Compelled by these facts, Einstein removed the
cosmological constant, stating that it was the \emph{biggest}
blunder of his life.

In 1964, the detection of CMBR by Penzias and Wilson \cite{penzias}
marked an important moment for observational cosmology, confirming
and stating the FLRW model as the standard model of cosmology. On
the other hand, improvement of observations, such as the
\emph{COsmic Background Explorer} (COBE) in 1992 \cite{cobe,cobe2},
and more recently, its successor, WMAP, has appointed some drawbacks
in the model as, for instance, the presence of the well-known
anisotropies of the CMBR, as well as the small temperature
fluctuations on large scales of the order of $\frac{\Delta T}{T}\sim
10^{-5}$ \cite{marter}, which cannot be explained rigourously by the
standard model without an adjustment of the mechanism
\cite{turner,aguiar}.

The development of observational experiments has reconsidered the
cosmological constant firmly in the last decade, which has revealed
that $\Lambda$ does not vanish with the effective value
$|\Lambda_{\;eff}|\sim 10^{-47}\;GeV^{4}\;$. A
non-vanishing cosmological constant is compatible with the ancient
globular clusters, reconciling with the matter density observed
\cite{krauss,krauss1,primack}. As we're going to point out in
subsection \ref{decc}, the value of the cosmological constant, if
regarded as a vacuum energy (as proved by the Casimir effect
\cite{casimir,mostepanenko,jaffe}), is at odds with the theoretical
value of the quantum energy density predicted by QFT. Indeed, the
situation has been aggravated by the decisive observations of the
accelerated expanding universe in 1998.

\subsection{The accelerated universe}
The first evidences of an accelerated expansion of the universe was
obtained from Hubble Space Telescope (HST) of current type Ia
supernovae (SNIa) in 1998 \cite{perlriess,perlriess2}, in agreement
with Chandra observations \cite{chandra}. Moreover, it was sustained
by measurements of the cosmic microwave background (CMB)
anisotropies \cite{bernardis} and the large scale structure data
\cite{spergel2}. The data suggested the existence of a density
energy component unclustered that fulfill 72$\%$ \cite{spergel} of
the universe with negative pressure driving the universe to an
accelerated phase of expansion, that is, a repulsive effect of
gravity. This effect was the so-called \emph{Dark Energy}, which is
corroborated by 250 independents astronomical observations events in
supernovae \cite{bertolami,schimidt}.

In principle, the first interpretation for the acceleration of the
universe is given by the FLRW standard model. According to
eq.\ref{eq:ray}, we can get two important conditions: first, the
\emph{strong} energy condition: when $(\rho+3p)>0$, then
$\ddot{a}<0$, i.e, the gravity decelerates the expansion of the
universe. Second, the \emph{weak} energy condition: when
$(\rho+3p)<0$, with negative pressure $p<0$, then gravity
accelerates the relative expansion between the structures of the
cosmos. This last condition is compatible with the experimental
observations of supernovae regarded as standard candles (stellar
objects that are used to infer distances based on its luminosity).
In a manner to try to deal with the observations, the cosmological
constant has been reconsidered as a dark energy component. The
presence of such term induces to an existence of a repulsive gravity
in universe and, in fact, turned out to be the most simple option to
deal with. The main debate now concentrates on the perturbation of
the Friedmann equations with $\Lambda$ as in eq.(\ref{eq:friedman}) or, 
besides the context of GR, $\Lambda$ plus a correction term
\begin{equation}\label{eq:friedman2}
\left(\frac{\dot{a}}{a}\right)^2+\frac{k}{a}=-\frac{8\pi G}{3}\rho+
\frac{\Lambda}{3}+\;(correction\;term)\;.
\end{equation}
In the following, we present some of proposals to solve the
dark energy dilemma.

\subsection{Dark Energy, the Cosmological Constant problem and some
alternatives}\label{decc} In fact, the dark energy problem is
related to the foundations of gravitational cosmological theory and
it has stimulated a demand for gravitational models and theories. In
this subsection,  amidst other proposals, we make general comments
of some current proposals to modify GR so it fits on the assumption
of extra-dimensional models. Most of these proposals try to solve
both dark energy and cosmological constant problems. Since the
cosmological constant can be regarded as dark energy candidate it is
inevitable the disassociation of them.

The so-called \emph{Cosmological Constant problem} had its first
seeds planted in 1916, with the ideas of Nernst \cite{nernst}. He
studied the non-vanishing vacuum energy density that was fulfilled
with radiation-only content, which was confirmed by the Casimir
effect in 1948 \cite{casimir,mostepanenko,jaffe}. Originally,
Casimir effect consists in the effect of approximation of two
separated uncharged conducting plates due to the zero-point energy
density of the electromagnetic field. The Casimir force is generated
by the energy density difference of pairs of virtual particles and
virtual-antiparticles between the plates and outside the plates.
Hence, the difference of pressure outside the plates are more
intensive than between the plates which generates a force on the
plates approximate them. In this manner, it was the first
experimental evidence of an existing non-vanishing contribution of
quantum vacuum energy density.

In late 1920's, Pauli \cite{pauli2,pauli3,rugh} made studies about
the gravitational influence of the vacuum energy density of the
radiation field, suggesting a conflict between the vacuum energy
density and gravitation. If vacuum energy density is considered,
then gravity must be dispensed. Moreover, based on Pauli's work,
Straumann \cite{pauli2,pauli3} restated that if one can consider the
static Einstein dust-dominated universe, the radius of the universe
would be of the order of 31km, lesser than the Earth-Moon distance,
thus confirming the conflicting Pauli's results that passed
unnoticed by scientific community. Even so, in the subsequent
decades, even with the dismissal of $\Lambda$ by Einstein, some
universe models based on $\Lambda$ were still studied, for instance,
the Lema$\hat{\mbox{i}}$tre model, as pointed out in subsection
\ref{ecosmo}. In addition, the observations of quasars in the
mid-late of the 1960's suggested the reconsideration of $\Lambda$
\cite{petrosian}.

\subsubsection{The quantum vacuum energy density as cosmological
constant} In 1967,  Zel'dovich \cite{zeldo} had a breakthrough
proposing the hypothesis in which $\Lambda$ is the vacuum energy. In
contrast with Pauli's conclusion,
 the vacuum energy density is considered, and gravity must also be
taken into account. By considering a perfect fluid, the vacuum
energy-momentum tensor $T^{\;vac}_{\mu\nu}$ can be given in comoving
coordinates as
\begin{equation}
T^{\;vac}_{\mu\nu}=\left(p_{\;vac}+\rho_{\;vac}\right)U_{\mu}U_{\nu}+p_{\;vac}g_{\mu\nu}\;\;,
\end{equation}
where $p_{\;vac}= - < \rho_{\;vac} >$ and $< \rho_{\;vac} >$ is the
expectation value of quantum vacuum energy density, which is the
analogue expression to the ordinary perfect fluid
$T^{\;mat}_{\mu\nu}$ in eq.(\ref{eq:fluid}). If we take Einstein
equations with $\Lambda$, we have
\begin{equation}
R_{\mu\nu}-\frac{1}{2}Rg_{\mu\nu}+\Lambda g_{\mu\nu}= 8\pi G
\left(T^{\;mat}_{\mu\nu}+T^{\;vac}_{\mu\nu}\right)\;\;.
\end{equation}
But, in vacuum, $T^{\;mat}_{\mu\nu}= 0$ and taking the covariant
derivative, we have
\begin{equation}
<\rho_{\;vac}>=\;constant\;\;,
\end{equation}
thus,
\begin{equation}
<\rho_{\;vac}>=\;\Lambda\;\;.
\end{equation}
Therefore, according to Zel'dovich $\Lambda$ can be regarded as a
quantum vacuum energy density when vacuum is regarded as a perfect
fluid, with $p=-\rho$. However, this situation would not reveal a
mere fact with the development of phenomenological observation
techniques. Effectively, the problem comes up because there is a gap
of the order of 123 decimal places between the cosmological observed
value of $\Lambda/8\pi G \approx 10^{-47}\; Gev^4$ and the
theoretical vacuum energy density prediction $< \rho_{\;vac}>\sim
10^{76}\;Gev^4$.

From a geometrical point of view, the cosmological constant problem is shown to be a
consequence of the \emph{equivalence class of metric geometries}
characterized by the Riemannian tensor. General relativity avoids
this difficulty by postulating the Minkowski space-time as the
standard flat geometry, from which we derive the concepts of
particles, quantum fields and their vacuum states. On the other
hand, the experimental evidences of a small but non-zero
cosmological constant is not compatible with the Minkowski
space-time, but it is consistent with the deSitter space-time.
Either we adopt the Minkowski flat-plane standard of curvature or,
in face of the observations we adopt the deSitter standard.
Therefore, it appears that the emergence of the cosmological
constant problem is a symptom of the lack of an independent
reference standard for curvature in Riemann's geometry.

On the other hand, as a realization of Heisenberg's uncertainty
principle, the theoretical value for the vacuum energy density is
obtained from the individual contribution of each oscillator of mass
\emph{m} and wave number $k_{max}$ cutoff in a set of harmonic
oscillators(with $\hbar=c=1$)
\begin{equation}\label{eq:vacuum}
<\rho_{\;vac}>= \int^{k_{max}}_{0} \frac{4\pi k^2dk}{2(2\pi)^3}
\sqrt{k^2+m^2}\sim \frac{k_{max}^4}{16\pi^2}\;.
\end{equation}
In order to avoid the ultra-violet(UV) divergence, we impose a
finite maximum value for $k_{max}$. And considering $\Lambda=(8\pi
G)^{1/2}$,  $<\rho_{\;vac}>$ results in $<\rho_{\;vac}>\sim
10^{76}\;Gev^4$, as stated before.

According to Weinberg \cite{weinberg}, the problem was taken
seriously in the 70's with the spontaneous breaking symmetry on the
electroweak scales. Even if we consider the lowest energy scale of
the quantum chromodynamics (QCD), which is of the order of
$0.3\;Gev$, we still have a huge difference of $46$ decimal places,
when compared to the cosmological observational value for $\Lambda$.
Such large difference cannot be eliminated by renormalization
techniques in quantum field theory unless an extreme fine tuning is
applied \cite{nobb,padman,weinberg}. 
But, why is the cosmological value of $\Lambda$ so small and can not
be regarded as zero? and why is it observed today? These are an
examples of unanswered questions. Even so, the cosmological constant
is one of the most important candidates to dark energy.

Therefore, the cosmological constant problem has become one of the
most important problems in modern physics, because is a problem of
fundamental nature, not only because it involves the structure of
the Einstein-Hilbert principle, but also because it apparently deals
with the distinction between gravitation and gauge fields.

\subsubsection{Some basic approaches about dark energy and cosmological constant}
In accordance with \cite{nobb}, we have chosen some
interpretations and proposals of solutions for the cosmological
constant problem and/or the dark energy problem. Here we point out
some models related to fine-tuning process, symmetry mechanisms,
violation of the \emph{equivalence principle} models and statistical
approaches. Most of these are extensively explored nowadays despite
of the lack of a proper explanation by first principle of a complete
theory.
\begin{flushleft}
\textbf{\small{Fine-tuning mechanisms}}
\end{flushleft}
As a \emph{fine-tuning mechanisms}, the simplest idea is to consider
the cosmological constant as a dark energy component, without
separating the concepts of quantum vacuum energy density and the
cosmological constant itself. As conceived in the
$\Lambda$-\emph{Cold Dark matter} ($\Lambda$CDM) model, the
cosmological constant is as a source term that obeys the cosmic
equation of state with pressure
$p_{\Lambda}=w_{\Lambda}\rho_{\Lambda}$, where $w_{\Lambda}=-1$ and
the energy density $\rho_{\Lambda}=\Lambda/8\pi G$. In spite of its
simplicity, it is in good agreement with experimental data from WMAP
and others astronomical data \cite{spergel,bennet}.

Another attempt was, rather than supposing $\Lambda$ as a constant,
regard it as a function of time. The time-varying ``cosmological
constant'' or $\Lambda(t)CDM$ predicts that the vacuum quantum
energy density decaying into CDM particles and transferring energy
to these particles. In fact, this model proposes the introduction of
a new field produced by $\Lambda(t)$ responsible for the
acceleration of the universe. The coupling between the
energy-momentum tensor $T_{\mu\nu}$ and $\Lambda(t)$ is a
consequence of the Bianchi identities when applied to Einstein's
equations and is given by
\begin{equation}
    T_{\mu\nu;\nu}= -\left(\frac{\Lambda(t)g_{\mu\nu}}{8\pi
    G}\right);\nu
\end{equation}
where the energy density and the cosmological scale factor $a(t)$
are related by the \emph{ansatz} $\rho_m=\rho_0a^{-3+\epsilon}$ by
taking a small deviation $\epsilon$ from the standard cosmic
evolution. And calculating the conservation of the energy-momentum
tensor $T_{\mu\nu;\;\nu}=0$ , one can find
\begin{equation}
    \dot{\rho}+3\frac{\dot{a}}{a}\rho=\dot{\rho_{\Lambda}}\;,
\end{equation}
where we denote the ordinary time derivative by a dot. And hence
\begin{equation}
    \rho_{\Lambda}=\bar{\rho_{\Lambda 0}}+ \frac{\epsilon
    \rho_0}{(3-\epsilon)}a^{-3+\epsilon}\;,
\end{equation}
where $\rho_{\Lambda}$ is the energy density contribution from
$\Lambda$, $\rho_0$ is the current CDM energy density and
$\bar{\rho_{\Lambda 0}}$ is an integration constant
\cite{ozer,alcaniz}.

In spite of some merits, the lack of some explanations from
$\Lambda$CDM to, for example, predict the cusped central density on
galactic sub-scales structures which are at odds with observations
\cite{salucci2,navarro,navarro2}, has motivated other phenomenological
proposals. One of them is the \emph{X-Cold Dark matter} (XCDM)
model, which is characterized by the equation of state
$p_{x}=w_{x}\rho_{x}$ of an exotic fluid. The $w_{x}$ parameter can
be a constant or more generally, a function of time
\cite{alcaniz2,alcaniz3,alcaniz4,alcaniz5}, which has a plethora of
proposals. The energy-conservation equation can be written as
\begin{equation}
\dot{\rho_{x}}=-3\frac{\dot{a}}{a}\rho_x(1+w_x)\;.
\end{equation}
Hence, if $w_x$ is a constant and $w_x<0$, one can find
\begin{equation}
\rho_x\sim a^{-3(1+w_x)},
\end{equation}
where $\rho_x$ is the density contribution from the \emph{X}-fluid.
One of the justifications for $w_x<0$ is that $w_x$ is a sufficient
smooth component, making it compatible with the age of the universe
as well as the rate of growth of the density perturbations in small
scales, plus it gives out solutions to the problems of redshift in
SNIa and gravitational lensing measurements \cite{turner}.

Moreover, the values of the parameter $w_x$ define different
cosmological scenarios. For instance, to reproduce an expanding
universe, we must set $w_x<-1/3$ which gives a large contribution to
$(\ddot{a}/a)$. And when $w_x=-1/3$, it imprints no effects on
$(\ddot{a}/a)$, i.e, it reproduces the same scenario as in the
standard open universe without any dark energy assumption. There is
still a weird scenario of a universe when $w_x<-1$ that proposes the
existence of some sort of exotic fluid that violates all energy
conditions and induces a huge increase of negative pressure, driving
the universe to a singularity at a finite time named \emph{Big-Rip},
where the factor scale and the curvature of the universe diverges
\cite{peebles,odintsov,caldwell2,alcaniz6,abramo}. This scenario is still an
odd possibility, since it was recently constrained by observations
of the Chandra x-ray observatory \cite{chandra}. For a flat universe
based on SNIa and CMBR data, we have $-1.11\leq w_x \leq
-0.86$\cite{cora,cora2}, and based on X-ray clusters and SNIa
surveys $w_x = 0.95^{+0.30}_{-0.35}$ \cite{Schu,Schu2,Schu3}. When
$w_x=-1$, we have the $\Lambda$CDM model, which fits to recent WMAP
observations \cite{nolta} on CBMR. Moreover, if we define
\begin{equation}
\Omega_x= \frac{8\pi G \rho_x}{3 H_0^2}\;.
\end{equation}
Assuming $w_x$ is constant and neglecting the current tiny
contribution of $\Omega_{R}$ and $\Omega_{k}$, we can rewrite
Friedman equation simply as
\begin{equation}\label{eq:xcdm}
H^2=H_0^2\left[\Omega_{M}(1+z)^3 +
\Omega_{x}(1+z)^{3(1+w)}\right]\;.
\end{equation}
To analyze the evolution the universe, we can study the deceleration
parameter, in terms of the redshift, given by
\begin{equation}
q(z)=\frac{1}{H}\frac{dH}{dz}(1+z)-1\;,
\end{equation}
where $H$ is given by eq.(\ref{eq:xcdm}). Thus, we obtain
\begin{equation}
q(z)=\frac{3}{2}\left[\frac{\Omega_{M}(1+z)^3 + (1+w)
\Omega_{x}(1+z)^{3(1+w)}}
{\Omega_{M}(1+z)^3+\Omega_{x}(1+z)^{3(1+w)}}\right]-1\;.
\end{equation}
and plot the behaviour of the deceleration parameter running the
values of $w$. The values $q= -0.6 \sim -0.7$ are the current values
for the deceleration parameter compatible with the constraints from
supernovae observations


There is another approach, where one can state that dark energy is
provided due to some sort of unclustered scalar field according to
the quintessence proposal in such a manner that drives the universe
to speed up. The \emph{quintessence} \cite{caldwell,peebles2} model
consists in an addition of a minimal coupled scalar field
$V(\varphi)$ to Einstein's equations, which yields a sought-after
extreme fine-tuning to solve the hierarchy discrepancy
\cite{waga,capo}. By writing the energy-tensor for a scalar field
$V(\varphi)$ as
\begin{equation}
    T_{\mu\nu}=\varphi_{,\mu}\varphi_{,\nu}-\left(\frac{1}{2}\varphi^{,\alpha}\varphi_{,\alpha}-V(\varphi)\right)g_{\mu\nu}\;.
\end{equation}
Using the conservation law for $T_{\mu\nu}$, one can find the
Klein-Gordon equation
\begin{equation}
    (\varphi^{;\mu}_{\;;\mu})+\frac{\partial V(\varphi)}{\partial
    \varphi}=0\;.
\end{equation}
Thus, the related field equation is
\begin{equation}
    \ddot{\varphi}+3H\varphi+\partial_{\varphi}V(\varphi)=0\;,
\end{equation}
where we denote $\partial_{\varphi}=(\partial/ \partial \varphi)$,
where $H$ is the Hubble parameter given by eq.\ref{eq:friedman} and
the total energy density is defined as $\rho=\rho_m +
\rho_{\varphi}$. The energy density of matter can be given by
eq.\ref{eq:conserv} while the energy density of the quintessence
field $\varphi$ is given by
\begin{equation}
    \rho_{\varphi}=\frac{1}{2}\dot{\varphi}^2+V(\varphi)\;.
\end{equation}
The underlying idea is to create a mechanism of decaying for the
energy of vacuum with a low varying expansion rate \cite{weinberg}.
In spite of consisting of a good scheme as a phenomenological model,
it lacks fundamental based grounds with \emph{ad hoc} proposal of a
quintessence potential. Nevertheless, some researches consider that
the quintessence field may not only be identified as the dark
component dominating the current cosmic evolution, but also as a
bridge between an underlying theory and the observable structure of
the universe \cite{alcaniz}.

\begin{flushleft}
\textbf{\small{Symmetry mechanism}}
\end{flushleft}
To proceed further, in the \emph{scale invariance approach} we have
unimodular theories \cite{dolgov} of gravitation as examples of
proposals.  Basically they modify Einstein-Hilbert principle
$S_{EH}$ by introducing a Lagrange multiplier $\mathcal{L}$ in a
manner to substitute the cosmological constant. This leads to a
modified Einstein-Hilbert $S_{mod\;EH}$ principle and a fixed
absolute space-time volume element, the so-called \emph{modulus}
\cite{finkelstein}. The Jacobian \emph{g} is regarded as a fixed
constraint, and as a result we have a new integration constant
$\Lambda'$ in the Riemann Scalar $R$ in such a way that
\begin{equation}\label{eq:modgrav}
S_{mod\;EH}=-\frac{1}{16 \pi G}\int d^4x
\sqrt{-g}\left(R-\mathcal{L}(g-1)\right)\;.
\end{equation}
By varying the modified action $S_{mod\;EH}$ with respect to the
metric $g_{\mu\nu}$, we can obtain
\begin{equation}
  R_{\mu\nu}-\frac{1}{4}Rg_{\mu\nu}=0\;,
\end{equation}
and using the Bianchi identities result in
\begin{equation}
    \partial_{\mu}R=0\;.
\end{equation}
Hence, $R$ is a constant and can be set as $R=-4\Lambda'$ which
allows us to write eq.(\ref{eq:modgrav}) as
\begin{equation}
  R_{\mu\nu}-\frac{1}{2}Rg_{\mu\nu}=\Lambda' g_{\mu\nu}\;.
\end{equation}
Although it has a reduction of 2 levels of degrees of freedom, due
to the constraints $R=-4\Lambda'$ and $det(g_{\mu\nu})\neq 1$,
$\Lambda'$ does not alter the dynamics of the equations and the
problem endures so that the fine-tuning mechanism is still
necessary. Consequently, $\Lambda'$ can take any value, even not
being explicitly provided in the modified Einstein-Hilbert action
\cite{silva2}. In spite of these shortcomings, the unimodular
theories have been studied and applied to cosmology nowadays
\cite{faraj}.

In addition, a more geometrical approach to the problem appeals
again to modify Einstein-Hilbert action principle, such as in the so
called $F(R)$ theories, using higher order Lagrangian, where the
higher order curvature terms provide for the difference between the
observed cosmological constant and the vacuum energy
\cite{odintsov2,capozziello}. However, once a physically justifiable
Lagrangian such as the Einstein-Hilbert principle is replaced by
$F(R)$, it becomes a necessity to properly justify a choice among a
large variety of options. More specifically, it appears that the present astrophysical observations are not sufficient to decide on what $F(R)$ to choose\cite{leszek}. On the other hand, recent
studies suggests that $F(R)$ cosmology provides a accelerating
behavior during attractor phase of matter-dominated era at odds with
expectations, with expansion factor varying as $a(t)\propto t^{1/2}$
\cite{amendola2}.

As we have discussed in section \ref{3}, in contrast with the dark
energy repulsive effect observed in cosmological scales, dark matter
is regarded as a sort of non-baryonic matter with merely attractive
effect. Besides, a local effect on small scales is suggested,
since it influences the growth of structures in the early universe.
Thus, at first, dark matter and dark energy constitute elements with
opposite gravitational characteristics and they are the main
characters of the cosmological ``tug-of-war'' \cite{dark}. However,
due to the lack of observational evidences, which can suggest that
theses components are generated by different sources, there are some
unification symmetry models that stick together both dark matter and
dark energy. A very known model in this approach is the quartessence
\cite{amendola}, which has as the main candidate some sort of exotic
gas called Chaplygin gas \cite{bilic}, with the equation of state
\begin{equation}
p= -\frac{A}{\rho}
\end{equation}
and, moreover, at a generalized form $p= -\frac{A}{\rho^{\alpha}}$,
where the parameters $A$ is restricted to $0< A < 1$ and $\alpha$ to
the range $-1< \alpha \leq 1$ in a manner of reproducing the early
and later times of the universe. Depending on the choice of
parameters, the gas behaves sometimes attractively and sometimes
repulsively, or equivalently, as similar to dark matter and dark
energy which could be related to some topological change in the
universe. However, it has some constraints \cite{spergel,makler}
based on SNIa experiments and statistics of gravitational lensing .

Another symmetry mechanism is related to supersymmetric models
where, for short, the cosmological problem does not occur.
Basically, a constraint imposed by supersymmetry (SUSY) on the
vacuum energy prevents it from even existing. Thus, the sum of the
contribution of al density states are canceled due to every
supersymmetric particle has an equivalent superpartner, hence,
$\Lambda= 0$. Even though $\Lambda$ does not exist in supersymmetric
models, it still is one of the most accepted proposals to explain
dark energy. The current explanation is that when supersymmetry is
broken, the dark energy, as conceived, comes up \cite{weinberg}, and
the scalar fields give the effective value of $\Lambda$ in order to
remedy this situation. Further information and recent works can be
found in \cite{cline, mairi, mairi2}.

\begin{flushleft}
\textbf{\small{Violating the equivalence principle}}
\end{flushleft}
The main example of violating the \emph{equivalence principle} of
General relativity is related to Brane-world models. Just like
Superstrings or M-theory, these models bring to light the discussion
of the existence of the extra-dimensions. They intend to provide a
solution to the hierarchy problem, and possibly a sought-after
unified physics theory of all conceived interactions.

In summary, these approaches are based on trying to decouple gravity
from vacuum energy density, making it indifferent to gravitation.
The general idea is that gravitation propagates in the
extra-dimensions in a sort of ``leakage'' of gravitation into the
bulk, in which the brane (4-dimensional space-time) is embedded in.
This hypothesis could explain how gravity is weaker than other
interactions as measured by an observer on the brane, where other
gauge interactions are confined. Actually, the confinement of the
gauge interactions has to do with special relativity, where the
standard model of particles and interactions are built on. In other
words, it is a consequence of the Poincar$\acute{\mbox{e}}$ symmetry
of the electromagnetic field, and in general, of the dualities of
the Yang-Mills fields, which are consistent in four-dimensional
space-time only. If we consider an odd brane-world quantum-gravity
theory the gravitational particles, named gravitons, are regarded as
some oscillations modes for the extra-dimensions.

One of the most known brane models is the Randall-Sundrum (RS) type
II model \cite{randall}. When applied to Cosmology, the vacuum
energy density in a 3-brane is smaller than the one predicted by
quantum field theory, which means that the cosmological constant
problem persists, even though the fundamental Tev scale energy is
preserved. A similar situation occurs when we treat dark energy
problem in which RS model II provides the modified Friedmann
equation
\begin{equation}
\left(\frac{\dot{a}}{a}\right)^2=\frac{8\pi}{3m_{pl}^2}\rho+
\frac{16\pi^2}{9m_5^6}\rho^2
\end{equation}
where $m_5$ is the 5-dimensional planck scale, $m_{pl}$ is the
4-dimensional planck scale. The correction term corresponds to the
square of the energy density $\rho^2$ of the confined matter
\cite{tujikawa,tujikawa2,maia3}. As it is well known, this result is
not compatible with recent observational data \cite{map,sdss} since
the additional term on Friedmann's equation, i.e, the energy density
$\rho^2$, provides a deceleration scenario of the universe, besides
affecting the nucleosynthesis of large structures. To remedy this
situation, other attempts have been studied, such as particular
classes of bulk and brane scalar potentials
\cite{kanti2} that lead to a fine-tuning mechanism.

Another proposal is the Dvali-Gabadadze-Porrati or DGP \cite{dvali}
model where the 5-dimensional bulk is flat and the brane is fixed,
that is, the embedding of the brane into the bulk is rigid with a
noncompact, infinite-volume extra dimension. It also presents some
difficulties related to strong interactions and massive gravitons
and it does not duly adjust to the accelerated expansion scenario,
even when studied on its general, the Dvali-Turner model
\cite{bertolami,nobb} which still requires, as it does the RS model,
an extreme fine-tuning to make it compatible with the observational
data. 

A promising brane-world approach stated in \cite{MaiaGDE} proposes a covariant (model independent) formulation of the brane-world theory based  on  the  perturbational  theory   of  local embedded
  submanifolds rather than particular junction conditions as commonly used in RS model and variants, hence the extrinsic curvature is considered as an independent field of spin-2 as compared with the metric. The main motivation of this approach has its roots in the classic problem in differential geometry,
originated in the early days of the Riemannian geometry, whose solution
was suggested by L. Schlaefli \cite{Schlaefli} in 1873, by comparing two geometries,
so that one is gauged by the other. The general solution for the
problem was given by J. Nash \cite{Nash} in 1956. Nash showed how
any Riemannian geometry can be generated by metric perturbations
against a bulk space (which he assumed to be Euclidean, but it was
soon extended to a pseudo Riemannian bulk by R. Greene \cite{Greene}). As it
happens, any embedded metric geometry can be generated by a
continuous sequence of small metric perturbations of a given
geometry with metric of the immersed manifold, i.e.,
\[
g_{\mu\nu}  =  \bar{g}_{\mu\nu}  +  \delta y \, \bar{k}_{\mu\nu}  +
(\delta y)^2\, \bar{g}^{\rho\sigma}
\bar{k}_{\mu\rho}\bar{k}_{\nu\sigma}\cdots\;\;.
\]
When applied to cosmology, the brane-world modified Friedman equation is obtained
 \begin{equation}
\dot{a}^2+k=\frac{8\pi G}{3}\rho a^2 +\frac{\Lambda}{3} a^2+
\frac{b^2}{a^2}\;\;, \label{eq:Friedmann}
\end{equation}
 where   the  $b(t)$ correction term with
respect  to the  standard  Friedman  equation is given by  the component $k_{11}(t)$
 of the extrinsic curvature. When compared with the x-fluid  state  equation
 $b(t)$ has the form
\begin{equation}\label{eq:b}
 b(t) =b_{0}(\frac{a}{a_{0}})^{\frac{1}{2}(1-3\omega_{x})}\;\;,
\end{equation}
where $a_{0}$   and   $b_{0}$   are  integration constants,
representing the  current expansion parameter and the current warp of the universe. 
From the theoretical point of view, it would be a satisfactory solution for the dark energy problem if the $b(t)$ function was a unique solution, but, in fact, it depends on a choice of a family of solutions for the extrinsic curvature induced by the homogeneity of the Codazzi equation which is well-known equation in differential geometry. Thus, to be  free   from these
  pathologies a proper mechanism or an additional dynamical equation for extrinsic curvature should be implemented. In spite of Brane-world models get some attention on recent
years due to several options for dark energy, their mechanisms are still
not completely understood or justified.

\begin{flushleft}
\textbf{\small{Statistical approach}}
\end{flushleft}
The \emph{statistical approaches} focuses basically on an
explanation for the value of the cosmological constant and, in a
general manner, the physical constants. The main debate concentrates
on why the physical constants have the value which are measured
today.

An example of statistical approach approach is the \emph{Anthropic
Principle} \cite{weinberg,carter}, mainly based on Bayesian
statistics. This principle plays a important role when applied to
Superstrings theory with an implementation of the Calabi-Yau
manifolds. These manifolds were used to explain the extra
6-dimensions of the theory built in 10 dimensions. For each one of
those compactions, there exists a wide landscape of possible
universes, where vacuum energy is anthropically allowed. Each one of
these universes have different values of $\Lambda$ and different
physics \cite{starkman}. This situation is solved with the strong
anthropic principle that states the universe we live in is the only
one adequate to man existence, rather, the law and physical
constants have the value they have to provide intelligent life. In a
weak version of the anthropic principle, Weinberg \cite{weinberg}
says that the intelligent life is the way it is only with the
minimum value oscillations of $\Lambda$ and $< \rho_{\;vac}>$ . 
On the other hand, in the strong version, the \emph{a priori} probability is more ``problematic'' because the \emph{ensemble} gathers the set of cosmological models fixed with different values of the fundamental physic constants violating the logical principle called \emph{Occam's razor} that states ``\emph{entia non sunt multiplicanda praeter necessitatem}'' (Entities should not be multiplied beyond necessity). Although interesting, such approach sets a too distant target for theoretical physics as an experimentally based discipline since we can only make experiments in one universe. Another criticisms and discussions of anthropic principle can be found
in \cite{garriga,smolin,aguirre}.

In addition, in a manner to avoid the anthropic principle, in the
superstrings context, the cosmological natural selection
\cite{smolin2}, by analogy to theoretical biology, gives an
explanation of the choices of parameters and fine-tuning process in
a landscape theory without appealing to the anthropic principle. The basic idea is that the universe was designated to the black-hole production stating an existing population of correlated universes. Such populations must attend to a very specific type able to evolve. For instance, the physical parameters are fixed in each universe but they can vary in different universes. If one of these universes can produce black-holes, it is called \emph{active universe}, i.e, in this universe \emph{child-universes} are produced in the event horizon of each black-hole. Each child-universe carries part of the characteristics (the values of the physical parameters) from the \emph{parent-universe}. The natural selection occurs precisely at the biggest possibility of an universe can dominate among an universe population, i.e, the biggest values of the physical parameters can be achieved in a manner of maximising the black-hole production, hence the child-universe birthrate and the rate of a life-permitting universes. Following this rationalisation, one can conclude that our universe with life is the result of an evolutive chain of birth and death of preceding universes. The main problems of this proposal are that there are not explicit reasons of why the choice of populations of specific characteristics evolve exactly by \emph{natural selection} and also if our universe is really the first universe or not. If the choice is \emph{randomly}, there is no \emph{progeny}, hence there is no natural selection \cite{maccabe}. In fact, the natural selection does not make an improvement upon the weak anthropic principle.

On the other hand, in the Horava-Witten's \cite{horava} superstring
model, the Calabi-Yau manifolds are not used. This model is built in
a 10-dimensional space-time, which is reducible to Anti-deSitter
$ADS_5$ space-time by using ADS/CFT correspondence, as proposed by
Maldacena \cite{maldacena} in 1998. By taking the Anti-deSitter
space-time in five dimensions, Maldacena concluded that every theory
built in the $M_4$ Minkowski space-time corresponds to a theory in
the $ADS_5$ space-time in which one can relate to the Yang-Mills
theory of a gravitational theory. Hence, the ADS/CFT correspondence
is extended to supersymmetry only in 10 dimensions. Thus,
Horava-Witten superstring model has gained more attention lately
because it does not appeal to the anthropic principle in a manner to
deal with the cosmological problem.

Moreover, when adding the holographic principle
\cite{nobb,batiz,batiz2,hooft,hooft2} to the M theory has brought up
some interesting questions. The holographic principle states that
all information contended to a physical system in a region of space
is defined by its surface and can be represented as a hologram in a
$ADS_5$ boundary in Horava-Witten's model. The hologram rules the
boundary regions in this same space in which contains at most one
degree of freedom per Planck area. This area is defined as a small
square with side
 $L_p$; that is the Planck length $L_p(10^{-33}\;
cm)$ \cite{damp}. Thus, the number of degrees of freedom that
describes such region is finite and much smaller than the one
expected on quantum field theory \cite{nobb} constrained by the
hologram mechanism. This fact could explain the smallness of the
cosmological vacuum energy density since the energy density
decreases with the area \cite{balazs}, considering that the universe
is large when compared to Planck scales. Such as the anthropic
principle, the hologram principle provides several discussions about
its validity and range, which only further observations can shed
light on these issues. For further information about the hologram
principle and dark energy see \cite{balazs,witten}.

Nevertheless, we note that there is a lack of a fundamental theory
which could give a satisfactory explanation to these fundamental
problems, and that also conciliates theory and phenomenological
data. To get to this final step, extra information about the
universe become more and more vital to unriddle the issues
surrounding the foundations of physics.

\section{Some recent and upcoming projects on Dark matter and Dark energy}
Despite of facing striking problems, the interest on topics such as
dark matter and dark energy, which are undoubtable related to the
inner foundations of physics, has increased. We live in a very
interesting moment of non-stop process of improvement of measurement
devices, like the space probes and the ground-based telescopes and
terrestrial laboratories around the world. One point to note is that
the consortium between several research centers and institutions all
over the world are the cornerstone to achieve success in all
scientific projects. Here, we restricted ourselves to short comments in a manner of giving some examples of current and
upcoming projects on modern cosmology and astrophysics related to
dark matter and dark energy.

\subsection{Dark matter experiments}
Due to the importance of the contribution of dark matter for the
total energy density of the universe, several experiments have been
proposed and carried out by many scientific groups around the world
in search of relic dark matter candidates.

The current experiments on dark matter focus mainly on \emph{Wimps}.
This apparent preference on \emph{Wimps} is due to the
characteristic of clumping of cold dark matter and, possibly, the
formation of a bulk around the galaxy. In this sense, the DArk
MAtter or DAMA collaboration is one of the projects focused on Wimps
detection, such as the solar axions, and perhaps a Simps detection
as proposed in ref.\cite{mitra}. The project is a result of a
initial collaboration between Italy and China, and \emph{a
posteriori} other groups from India, UK, Russia, Ukraine and Spain
\cite{dama}. The method is based on the measurement of the annual
modulation signature which accounts for the supposed variation of
signal of dark matter due to the positions and velocities of the
Earth and Sun with respect to the galactic plane \cite{khalil}; the
material used is NaI crystal scintillator detectors. The initial
results of measurements were controversial due to the negative
results from other experiments, as, for instance, the US project
called the Cryogenic Dark Matter Search (CDMS)
\cite{akerib,akerib2}. Current efforts are being made in a manner to
reconcile DAMA experiments with other projects in order to gain
maximum level of confidence on measurements. A more detailed
information about the project and recent results can be found in \cite{bednyakov, dama, petriello, bernabei}.

To proceed further, The Cern Axion Solar Telescope (CAST)
\cite{cast, zioutas} a collaboration between Germany, Greece, Italy,
UK, USA and Russia, is intended to detect axions that escape from
the solar core. The CAST apparatus is basically composed of a Large
Hadron Collider (LHC) prototype magnet of order of $9T$. It resides
in the interior of two parallel pipes of length $L=9,26m$, and
cross-sectional area $A=2\times14,5 cm^2$. With the current upgrade
of LHC it has been expected information about, for example, dark
matter, production of mini-black holes, the existence or not of
extradimensions and the Higgs boson, and possibly the appearance of
new paradigms in physics. Due to the high level of sensitivity, as
stated in \cite{horvat}, CAST experiment can provide a probe for
the existence of extra-dimensions. The first phase results reached
the limit bound of axion mass $m_a $ of order of 0.20eV. The
on-going second phase intends to reach a region mass of order of
1eV, which can provide the observation of some new effect
\cite{adri}. Moreover, the Tokyo axion helioscope is a Japanese
experiment which is based on ``helioscope method'' which intends to
reach the same limit mass bound as the CAST experiment (1Gev) on its
current third phase. The experiment has been applied extensively on
the study of solar axions since 1997, and the actual status quo is
the on-going third phase. The apparatus is composed basically of a
2.3-m long 4T superconducting magnet, a gas container (for hydrogen
or helium), PIN-photodiode X-ray detectors, and a telescope. For
further information see \cite{yonue,yonue2}.

Another interesting experiment is the Cryogenic Rare Event Search
with Superconducting Thermometers (CRESST) \cite{cresst, cresst2}.
It is based on the analysis of the elastic scattering of the relic
particles. The possible small energy of the recoil can be detected
by sensitive cryogenic detectors. Just like DAMA experiment, the
CRESST apparatus is located at the Gran Sasso National Laboratory,
about 1.4 km below ground. Both experiments have such caution to
avoid an interference of any kind. The project has been updated to
its second phase started in 2007 and it has expanded
cryogenic detectors with scintillating crystals up to 33 detector
modules \cite{cresst}.

Several other experiments on dark matter have been proposed and
explored, SOLAX \cite{solax} and COSME \cite{cosme} to name a few.
All of these are in search of the relic particles, but none have
conclusive facts yet. Nevertheless, the efforts are in progress and
more sensitive devices have been constructed for detecting such
particles. Recently, physicists of the international collaboration
DZero experiment at the U.S. Department of Energy's Fermi National
Accelerator Laboratory have discovered ``doubly strange''
\cite{abazov} particle called the Omega-sub-b $(\Omega_b^{-})$ which
is constituted with two strange quarks and a bottom quark. However,
a profound analysis must be applied to study and understand this new
discover. A more complete list of dark matter experiments can be
found in \cite{list}.

\subsection{Projects on dark energy}

Just like dark matter experiments, the dark energy surveys have been
extensively explored in the recent years, mainly on space probes and
the ground-based telescopes. One very known example of a space probe
is the WMAP experiment \cite{map} launched in 2001, intended to
measure the CMBR anisotropies. The CMBR collected data plays an
essential role in modeling and analyzing models of theories about
the universe. The project is a partnership between Princeton
University and NASA's Goddard Space Flight Center. As stated in
former section, the current fifth year has shown us an improvement
of the measurement of the power spectrum and the composition of the
universe. In fact, the WMAP has been considered an essential tool
for the current cosmology and astrophysics.

In contribution to calibrate with WMAP collected data, we can
mention the CBI and BOOME-\newline RANG experiments. The Cosmic
Background Imager (CBI) is a radio telescope designed to study the
CBMR fluctuations on arcminute scales at frequencies between 26 and
36 GHz. It is located in the Chilean Andes, at the Chajnantor
Observatory. This project is a collaboration between the California
Institute of Technology, the Canadian Institute for Theoretical
Astrophysics, the University of Chicago, the National Radio
Astronomy Observatory, the Max-Planck-Institute f$\ddot{\mbox{u}}$r
Radioastronomie (Bonn), Oxford University, the University of
Manchester, the Universidad de Chile, and the Universidad de
Concepci$\acute{\mbox{o}}$n \cite{cbi,cbi2}. Moreover, the BOOMERANG
balloon, just like WMAP and CBI, is also designed to measure
anisotropies in the CMB. It consists of an array of detectors cooled
to 0.28 Kelvin mounted at the focus of a 1.3 meter telescope. The
instrument flows on a gondola beneath a NASA/NSBF high altitude
balloon \cite{boomerang}.

In addition, the PLANCK space probe \cite{planck} programmed from
the European Space Agency is designed to measure the anisotropies of
the CBMR in a manner to provide more information, besides sharpening
recent data about the universe. It also tests and makes constraints
on cosmological models and theories. The space probe has basically a
1.5 meter off-axis telescope which will be capable of covering the
entire sky with a precision of approximately 2 parts per million. It
is planned to be launched on October 2008, the Operations are on
2009-2010, and the scientific product delivery, is dated to mid
2012.

Besides the space probes and balloons, there are important
contributions from the grounded-based telescopes. They play a
fundamental role in providing reliable data surveys comparing to
other surveys, helping us to trail with higher level of confidence
the collected data. As it happens, we point out the spectroscopic
project named Sloan Digital Sky Survey (SDSS) \cite{sdss}, which
plays a fundamental role in providing data for dark energy issues.
It is essentially a 2.5 meter spectroscopic telescope located on
Apache point. Its first and second-finished phases of operations
were active during 2000-2005 and 2005-2008, respectively. In
addition, the successor of the SDSS-II, the SDSS-III is the current
research program and it is planned to start operating in june 2008
until 2014 with four new surveys using SDSS facilities.

In five years of operations, the SDSS program collected
approximately 200 million cosmic objects, such as galaxies, quasars
and stars. As the result of a consortium between 25 institutions,
the SDSS-II carried out 3 fronts of surveys: the Sloan Legacy, which
was applied to make a general survey of the cosmos; the Sloan
Extension for Galactic Understanding and Exploration (SEGUE
experiment), which was driven specifically to study the Milk-way,
and the Sloan Supernovae project, which was turned to study and
detect supernovae in the universe. This project will be improved
with the new surveys driven by the SDSS-III program: the Baryon
Oscillation Spectroscopic Survey (BOSS) will be intended to measure
the cosmic distance scale and mapping luminous distant galaxies and
quasars. The SEGUE-2 experiment is a second phase of the SEGUE
experiment started by the SDSS-II program and it will be intended to
make an improvement of studying of the Milky way structure,
kinematics, and chemical evolution. The study of inner structure of
the Milky way will be contemplated by the third program called the
APO Galactic Evolution Experiment (APOGEE) based on high-resolution
infrared spectroscopy. The last survey is the Multi-object APO
Radial Velocity Exoplanet Large-area Survey (MARVELS) intend to
locate and study distant bright stars and possible giant extrasolar
planets. In this manner, the SDSS program is intended to cover a
surprisingly amount of data and range of several scales of the
universe.

Another project is the upcoming Dark Energy Survey project(DES),
expected to start operating in September 2009, during 5
observational seasons till 2014. It is an international
collaboration between 10 institutions (at present time), and is
managed by Fermilab, The University of Illinois' National Center for
Supercomputing Applications (NCSA) and The National Optical
Astronomy Observatory (NOAO) \cite{des}. The DES project is designed
to study dark energy and its influence on the universe. It will use
a device for photometric surveys constituted of a 62 CCDs camera
(approximately 500 megapixels) coupled to  the 4-meter Blanco
telescope on Cerro Tololo Inter-American Observatory (CTIO) in
Chile. With 3 square-degree-field it is capable of covering an
eightvo of the sky in 4 bandpasses and it will detect approximately
300 million cosmic objects improving at least, for example, the SLSS
surveys by a factor 2. It is a powerful experiment on the dark
energy issues, and it will give us a large amount of collecting data
in the next couple of years.

Truth of matter, all the former projects will indirectly provide
ground to the ambitious Large Synoptic Survey Telescope (LSST). This
project will consist of a ground-based 8.4-meter and 10
square-degree-field telescope, schedule to operate in 2014, and it
is a conjoined effort between nineteen other organizations atop
Cerro Pachon in Chile. Due to its advanced hardware devices, it will
be capable of covering the sky every three nights. This project
certainly will imprint a massive impact on the observational
astrophysics and Cosmology ever seen with its great power of
resolution and amount of data, providing over 3 Gigapixels per image
to be processed and study. For further information see \cite{lsst}.

No other moment in mankind's history we have had so much
information about the universe, about its content and evolution,
even if our understanding about it is unsatisfactory due to the lack
of consistent and general explanations of the phenomena. Clearly,
all these host of recent and upcoming experiments will help us to
shed light on the \emph{dark} problems and issues of physics. To
better summarize the importance of these projects on current issues
we quote Gary Hinshaw of NASA's Goddard Space Flight Center in
Greenbelt \cite{map}
\begin{center}
``\emph{Ours is the first generation in human history to make such
detailed and far-reaching measurements of our universe}''.
\end{center}

\section{Concluding remarks}
Modern Cosmology has been an important source of data that provides
a deeper comprehension of the gravitational structure and evolution
of the universe. Not only this, but it calls for new gravitational
theories far beyond Einstein's approach. Even though we are long way
from a concrete fully-developed theory, dark matter and dark energy
play a major role on this quest, representing fundamental
constraints to these new gravitational models.

To stay on the right track, we need to use the information we have
at hands to make the best possible decisions. Based on recent data
in Astrophysics, mainly on the WMAP experiment and SDSS surveys, we
notice that dark energy is a disturbed element on the geometry of
the universe which is best described by the
Friedmann-Lema$\hat{\mbox{i}}$tre-Robertson-Walker model. The same
characteristic seems to appear on primordial dark matter issues,
which are related to the formation of large structures in the
universe. It is also important to point out that the Cosmological
Constant problem is keenly related to the Dark Energy problem, which
means that not only we have to understand dark matter and dark
energy better, but also its relation with the Cosmological Constant,
 which is still
itself the only viable explanation to conciliate theory and
phenomenological data. More recently, the problem has escalated to a
central issue in the context of the $\Lambda$CDM cosmological
scheme, where the accelerated expansion of the universe is
explained. Although the problem is rooted in General Relativity, it
shows up in other branches, such as non-supersymmetric theories,
like brane-world gravity. Moreover, the cosmological constant
problem can be taken as a problem of fundamental nature because it
involves the Einstein-Hilbert principle, and also because it
involves the different nature and energy scales of the four
fundamental interactions. Hence, we need a theory to deal with these
intrinsic features of data phenomenology, and not just a simple
mechanism, such as a fine-tuning, but indeed, a reasonable
description of the first principles.

As it is laid out right now, from a theoretical point of view, the
Cosmological Constant problem is a fundamental problem, such as the
hierarchy problem, and its solution must come from a complete theory
independent of particular models. As we want to point out to the
reader, besides a \emph{quantitative} difference between gravity and
gauge interactions, gravity \emph{seems} to behave
\emph{qualitatively} different than other gauge interactions, as we
see on the efforts of unifying all fundamental interactions, as well
as dark component problems and observational data. Clearly, the
current examples of gravitational effects imprinted by dark matter
and dark energy require a deeper insight on the structure of the
universe and on what gravity really means. In truth, these are
problems of fundamental nature that exercise a serious effect on the
foundations of gravitation conceived since the studies of Galileo,
Newton and Einstein. 

\section*{Acknowledgments} We would like to thank to A.R. Queiroz for the
criticisms and profitable discussions about this work and CNPQ of
Brazil for the funding support.

\end{document}